\documentclass[11pt]{article}

\usepackage{geometry}
\usepackage[toc,page]{appendix}

\usepackage{verbatim}
\usepackage{bbm}
\usepackage{amssymb}
\usepackage{amsmath}
\usepackage{amsthm}
\usepackage{dsfont}
\usepackage[dvipsnames]{xcolor}
\usepackage{ stmaryrd }
\usepackage{ bbold }

\usepackage{titlesec}
\titlelabel{\thetitle.\quad} 

\titleformat{\paragraph}
{\normalfont\normalsize\bfseries}{\theparagraph}{1em}{}
\titlespacing*{\paragraph}
{0pt}{3.25ex plus 1ex minus .2ex}{1.5ex plus .2ex}

\usepackage{graphicx}

\usepackage{tikz}
\usepackage{pgfplotstable}
\usepackage{pgfplots}
\usepackage[nomessages]{fp}
\usepackage{ifthen}

\usetikzlibrary{external}
\tikzexternalenable

\usepackage{algorithm,algpseudocode}

\usepackage[colorlinks = true, linkcolor = blue,
            urlcolor  = blue,
            citecolor = blue,
            anchorcolor = blue,
            pdfborder={0 0 0},
             plainpages=false,pdfpagelabels]{hyperref}

\usepackage[round]{natbib}
\newcommand{\mn }{\bar{m}}  

\newcommand{\dd }{m^{\prime}}

\newcommand{\diffb}{\triangledown}
\newcommand{\difff}{  \vartriangle \hspace{-0.8mm}}

\newcommand{\sbt}{ \hspace{0.2mm} \begin{tikzpicture} 
\filldraw[fill=black,draw=black] circle (0.6pt);
\end{tikzpicture}
}

\makeatletter
\newcommand{\oset}[3][0ex]{%
  \mathrel{\mathop{#3}\limits^{
    \vbox to#1{\kern-2\ex@
    \hbox{$\scriptstyle#2$}\vss}}}}
\makeatother

\newtheorem{defn}{Definition}

\newtheorem{exemp}{Example}

\newtheorem{propr}{Proposition}

\newtheorem{remark}{Remark}

\usepackage{hyperref}

\begin{document}

{
   \title{\textbf{Changepoint detection in non-exchangeable data}}
 \author{Karl L. Hallgren$^{1}$\thanks{Email: karl.hallgren17@imperial.ac.uk}, Nicholas A. Heard$^1$ and Niall M. Adams$^1$\vspace{.5cm} \\
    $^1$Department of Mathematics, Imperial College London
}
\date{\vspace{-5ex}}
\maketitle
}

\begin{abstract}
Changepoint models typically assume the data within each segment are independent and identically distributed conditional on some  parameters which change across segments. This construction
may be inadequate when data are subject to local correlation patterns, often resulting in many more changepoints fitted than preferable. This article proposes a  Bayesian changepoint model  which relaxes the assumption of exchangeability within segments. 
The proposed model supposes data within a segment are $m$-dependent for some unkown $m \geqslant0$ which may vary between segments, resulting in a model suitable for detecting clear discontinuities in data which are subject to different local temporal correlations. 
The approach is suited to both continuous and discrete data. 
A novel reversible jump MCMC algorithm is proposed to sample from the model; in particular, a detailed analysis of the parameter space is exploited to build proposals for the orders of dependence. 
Two applications demonstrate the benefits of the proposed  model: computer network monitoring via change detection in count data, and segmentation of financial time series.
\end{abstract}


\section{Introduction}

Standard changepoint models rely on partitioning the passage of time into segments, and fitting relatively simple models within each segment. In particular, the data within each segment are often assumed to be independent and identically distributed conditional on some segment specific parameter \citep{Green1995, exactcpt, Fryzlewicz}. This construction assumes the data within each segment are exchangeable, rendering the order in which the data are observed irrelevant when calculating their joint likelihood \citep{BernardoSmith}. There are many examples of applications where this assumption 
is suitable; see for example \citet{olshen}, \citet{Fryzlewicz} and \citet{fearnheadOutliers} for the detection of changes in the mean or variance of time series.

\begin{figure}[t!]
\vspace{-20mm}
\centering
\includegraphics[height=0.75\textwidth]{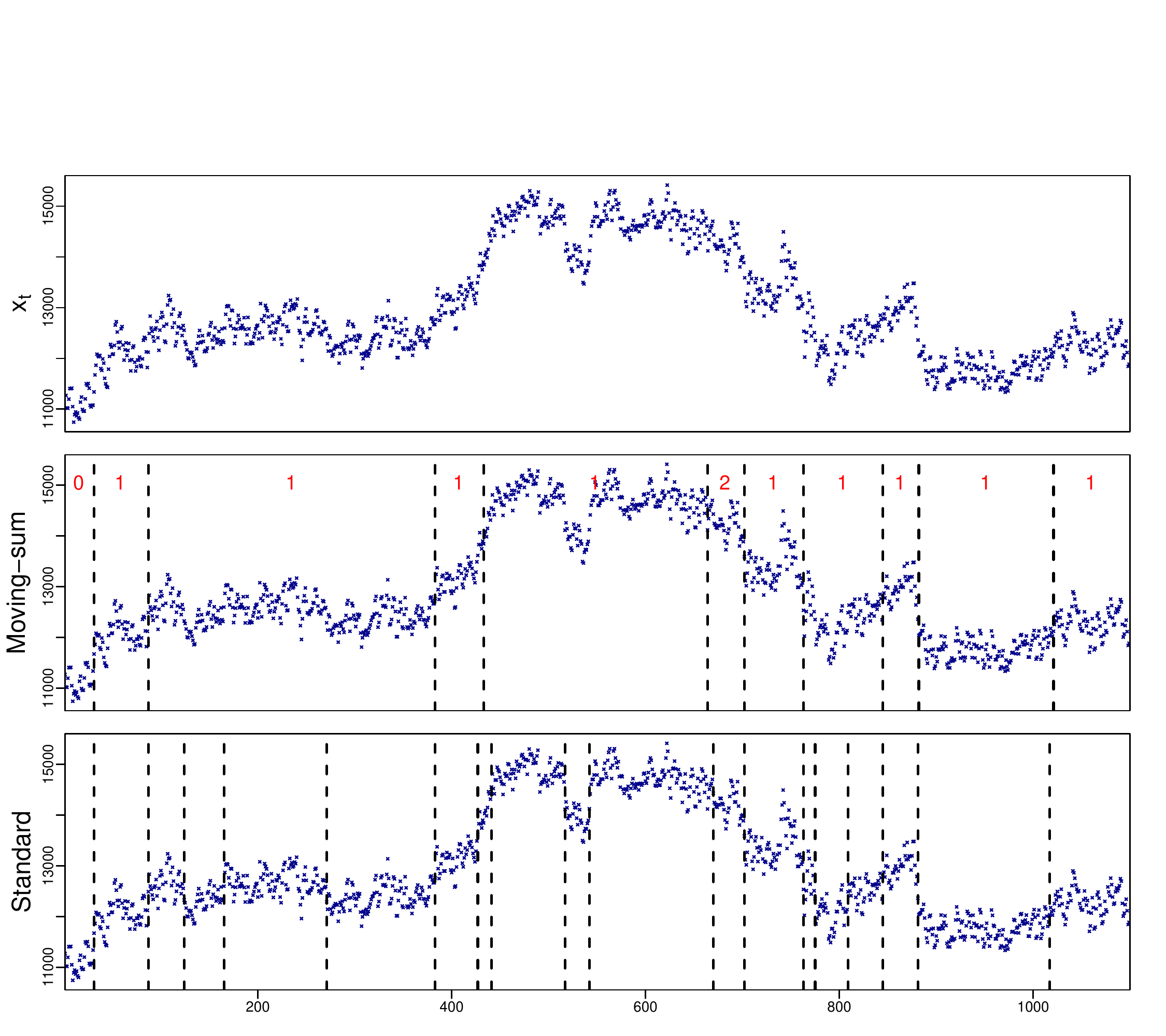}
\caption{Counts of computer network events each second between 10:00 and 10:20 on day $22$ of the data collection period.  Vertical lines indicate estimated changepoints for the proposed changepoint model (middle panel) and for the standard changepoint model \citep{exactcpt} (bottom panel). Numbers in red indicate the MAP order of dependence $m$ for each segment for the moving-sum model. }
\label{fig:Netflow}
\end{figure}

However, for some applications it can be reductive to assume  data are exchangeable within segments. 
For illustration purposes, we consider an application of changepoint detection in  
computer network monitoring. A cyber-attack typically changes the behaviour of the target network. 
Therefore, to detect the presence of a network intrusion, it might be informative to monitor for changes in the volumes of different types of traffic passing through a network over time. Yet, cyber data are often subject to 
population drifts, seasonal variations and other temporal trends that are unlikely to be evidence for cyber-attacks. As a result, traditional changepoint detection methods, which assume the data are exchangeable within segments, will fail to capture temporal dynamics and consequently fit many more changepoints than necessary.
For example, consider the second by second counts of network events recorded on the Los Alamos National Laboratory enterprise network \citep{unifieddata} which are displayed in Figure \ref{fig:Netflow}; the data will be presented in more detail in Section \ref{sec:cyberapp}. The vertical lines in the bottom plot indicate the maximum a posteriori (MAP) changepoints obtained with the  standard changepoint model described in \citet{exactcpt}, which assumes data are exchangeable within segments. More changepoints are fitted than preferable for anomaly detection. It would be desirable for abrupt changes, such as the ones observed near the 420th and 880th seconds, to be detected, whereas changes due to small fluctuations, such as the ones between the 100th and 380th seconds, may not be relevant to a cyber-analyst. 
A changepoint model to detect clear discontinuities in the presence of non-exchangeable data is needed. Moreover, since temporal dynamics for cyber data may change when a clear discontinuity occurs, it would not be satisfactory to assume the dependence structure of the data is the same for each segment.

Existing models to detect clear discontinuities in the presence of non-exchangeable data typically assume the dependence structure is identical for each segment \citep{ARchangesChib, Spark, Chakar, DeCAFS}. In particular, it is often assumed  the data within each segment are Markov conditional on some segment parameter. Moreover, changepoint models for dependent data are often designed for a specific marginal distribution, for example the normal distribution \citep{Chakar, DeCAFS}, the negative binomial distribution \citep{Spark, Yu2013} or the Poisson distribution  \citep{Weiss, Franke}.

This article extends a standard changepoint model \citep{exactcpt}, relaxing the assumption that data are exchangeable within segments. The proposed changepoint model, named the moving-sum changepoint model, supposes a segment model that is 
related to a model for $m$-dependent, stationary data discussed in \citet{Joe}; a sequence $x_1, x_2, \ldots$ is $m$-dependent if  $(x_{t+m+1}, x_{t+m+2}, \ldots)$ is unconditionally independent of $(x_1, x_2, \ldots, x_t)$ for all $t \geqslant 1$.
Within each segment, our model assumes the data are $m$-dependent and identically distributed conditional on some parameter $\theta$, where both $\theta$ and $m \geqslant 0$ are unknown and change from one segment to the next. Whilst $\theta$ denotes a parameter of the marginal distribution of the data such as the mean or the variance, $m$ corresponds to the level of dependency within the segment. 
To maintain tractability, the marginal distribution of the observed data is assumed to belong to the class of convolution-closed infinitely divisible distributions, which includes, for example, the normal, negative binomial and Poisson distributions. Therefore, the moving-sum changepoint model is  suitable for various settings where it is of interest to detect clear discontinuities in the presence of non-exchangeable data. 
For example, consider the MAP changepoints obtained with the moving-sum model for the counts of network events displayed in the middle panel of Figure \ref{fig:Netflow}. In comparison with the standard changepoint model, the moving-sum changepoint model captures temporal dynamics of the network behaviour, resulting in a segmentation of the data that is more adapted to network monitoring. 

A common approach to sampling changepoints for a time series is that of \citet{Green1995}, using a reversible jump MCMC algorithm to explore the state space of changepoints. At each iteration of the algorithm, one of the following move types is proposed: 
sample a segment parameter,  propose a new changepoint, or 
 delete or shift an existing changepoint to a new position. 
This article proposes a sampling strategy within that framework to sample from the moving-sum changepoint model. 
In particular, our approach exploits an analysis of the constraints of the parameter space; when the support of the data is non-negative, the constraints of the parameter space depend on the observed data, and this must be understood to build proposals for segment specific dependency levels.

The remainder of the article is organised as follows. 
Section \ref{sec:movingsummodelfull} introduces a novel changepoint model for non-exchangeable data. 
Section \ref{sec:conditionallik} gives an approach for deriving the likelihood of the data conditional on proposed changepoints, characterising the segment model in terms of a stochastic difference equation with an unknown initial condition. 
Section \ref{sec:parameterspace} provides a detailed analysis of the constraints to the parameter space, along with asymptotic results on the behaviour of segment parameters. A reversible jump MCMC sampling strategy is given in Section \ref{sec:rjmcmc}. Section \ref{sec:simulatoinstudy} presents results demonstrating the benefits of the proposed changepoint model, via a comparison with the standard model \citep{exactcpt} and DeCAFS \citep{DeCAFS}. Section \ref{sec:applications} considers two applications of changepoint detection showing the benefits of the proposed changepoint model: computer network monitoring via change detection in count data, and detection of breaks in daily prices of a stock.

\section{Moving-sum changepoint model}
\label{sec:movingsummodelfull}
This section introduces the moving-sum model, which is used as a segment model to define a novel changepoint model for non-exchangeable data.

\subsection{Moving-sum model}
\label{sec:movingsumsegment}
A moving-sum model assumes that observed data $x_1, \ldots, x_n$ satisfy
\begin{align}
\label{eq:movingsummodel}
x_t \overset{}{=} \sum_{i=0}^{m}y_{t-i},
\end{align}
for $t=1, \ldots, n$, where $y_{-(m-1)}, \ldots, y_n$ are $m+n$ iid latent random variables with common parametric density $f_m( \cdot \,  | \, \theta)$ for some unknown  parameters $\theta \in \Theta$ and $m \geqslant 0$. 
If $m=0$, the construction in (\ref{eq:movingsummodel}) implies that, for all $t=1, \ldots, n$,
\begin{align}
\label{eq:classicconiid}
x_t \overset{\text{iid}}{\sim} f_0( \cdot \, | \, \theta), 
\end{align} 
and, consequently, is equivalent to exchangeability in the data. 
Yet, if $m>0$, the sequence of observed data (\ref{eq:movingsummodel}) is $m$-dependent and therefore non-exchangeable.
\begin{defn}[$m$-dependence]
\label{def:markovdep}
For $m \geqslant 0$, the sequence $x_1, x_2, \ldots$ is $\, m$-dependent if $(x_{t+m+1}, x_{t+m+2}, \ldots)$ is unconditionally independent of $(x_1, x_2, \ldots, x_t)$ for all $t \geqslant 1$. Note that if a sequence is $m$-dependent, then it is also Markov of order $m$. 
\end{defn}
For all $t$,  $x_t$ in  (\ref{eq:movingsummodel})  is the sum of $m+1$ latent random variables, leading to $m$-dependence. Noting this duality, for simplicity of presentation in the following discussion we use the notational convention 
\begin{equation}
\bar{m} = m+1.
\end{equation}

It will be helpful to identify a class of distributions for which the construction in (\ref{eq:movingsummodel}) gives rise to a tractable marginal distribution of the observed data. Recall the distribution of a random variable $x$ is \textit{infinitely divisible} if, for all $m \geqslant0$, there exists a sequence of iid random variables $y_0, \ldots, y_m$ such that $\sum_{i=0}^{m}y_i$  has the same distribution as $x$. For all infinitely divisible marginal distributions $F$ for $x_t$ (\ref{eq:movingsummodel}), there exists a distribution $F_m$ for the latent random variables for all $m$, and $F_m$ is known if $F$ is closed under convolution. In this article, it will be assumed that the marginal distribution of $x_t$ is an infinitely divisible distribution which is closed under convolution, so that the corresponding density $f_m( \cdot \, | \, \theta)$ of the iid latent variables is available for all $m$.

We consider in detail three instances of the moving-sum segment model based on such distributions, one for continuous data with unbounded support given in Example \ref{eg:normalgamma}, one for continuous data with bounded support given in Example \ref{eg:gammagamma}, and one for discrete data with bounded support given in Example \ref{eg:negbin}, which we will refer back to throughout the article for illustration. 

\begin{exemp}[Normal marginal distribution]
\label{eg:normalgamma}
Suppose that $f_m( \cdot \, | \, \theta)$ corresponds to density of the normal distribution with mean $ \mu / \bar{m}$ and variance $\sigma^2 / \bar{m}$, for some $\theta = ( \mu, \sigma)$ where $\mu  \in \mathbb{R}$ and $\sigma>0$. It follows that $(x_t)$ is marginally 
$N(\mu, \sigma^2)$ with $m$-dependence. Moreover, a priori $\sigma^{-2} \sim \text{Gamma}(\alpha, \beta)$, for some $\alpha >0$ and  $\beta>0$, and $\mu \sim N(\mu_0, \sigma^2 / \lambda )$ for some $\mu_0 \in \mathbb{R}$ and $\lambda > 0$. 
\end{exemp}
\begin{exemp}[Gamma marginal distribution]
\label{eg:gammagamma}
Suppose that $f_m( \cdot \, | \, \theta)$ corresponds to density of the gamma distribution with shape parameter $\lambda/\bar{m}$ and rate $\theta$, where $\lambda>0$ and $\theta>0$. It follows that $(x_t)$ is marginally
$\Gamma(\lambda, \theta)$ with $m$-dependence. The prior for $\theta$ is assumed to be $\Gamma(\alpha, \beta)$ for some $\alpha>0$ and $\beta>0$.
\end{exemp}
\begin{exemp}[Negative binomial marginal distribution]
\label{eg:negbin}
Suppose that $f_m( \cdot \, | \, \theta)$ corresponds to density of the negative binomial distribution with number of failures $ r / \bar{m}$ and success probability $\theta \in [0, 1]$, for some fixed $r>0$.  It follows that $(x_t)$ is marginally $NB(r, \theta)$ with $m$-dependence. Moreover, 
a priori $\theta \sim \text{Beta}(\alpha, \beta)$, for some $\alpha >0$ and  $\beta>0$.
\end{exemp}

Other examples of such distributions include the Poisson, Cauchy and chi-squared distributions.

\subsection{Bayesian changepoint analysis with moving-sums}
\label{sec:fullchangepointmodel}
Suppose we observe real-valued discrete time data $x_{1:T} = (x_{1}, \ldots, x_{T})$. 
The changepoint model assumes $k \geqslant 0 $ changepoints with ordered positions $\tau_{1:k} = (\tau_1, \ldots, \tau_k)$, such that $1 \equiv \tau_{0} < \tau_{1} < \cdots < \tau_{k} <  \tau_{k+1} \equiv T+1$, which partition the passage of time into $k + 1$ independent segments. The changepoints are assumed to follow a Bernoulli process, implying a joint prior probability mass function $\pi( k, \tau_{1:k} ) = p^{k}(1-p)^{T-1-k}$ for some $0<p<1$.

For the moving-sum changepoint model, within each segment $j$, the data $x_{\tau_{ j-1}}, \ldots, x_{\tau_{j}-1}$  are assumed to follow the moving-sum model (\ref{eq:movingsummodel}) conditional on some unknown dependency level $m_j\geqslant 0$ and parameter $\theta_j\in \Theta$, which both change from one segment to the next. 
Dependency levels $m_1, \ldots, m_{k+1}$  and segment parameters $\theta_1, \ldots, \theta_{k+1}$ are assumed to be independent. It is assumed a priori that $m_j$ is drawn from a geometric distribution with parameter $0< \rho<1$, meaning the order of dependence may be increased or decreased at a fixed cost for each segment. Moreover, motivated by computational considerations, the prior for $\theta_j$ is chosen to be conjugate for $f_0$. 
For notational simplicity, we denote by $\pi$ the density of the prior distribution of both $m_j$ and $\theta_j$.

\begin{figure}[t!]
\centering
\includegraphics[height=0.36\textwidth]{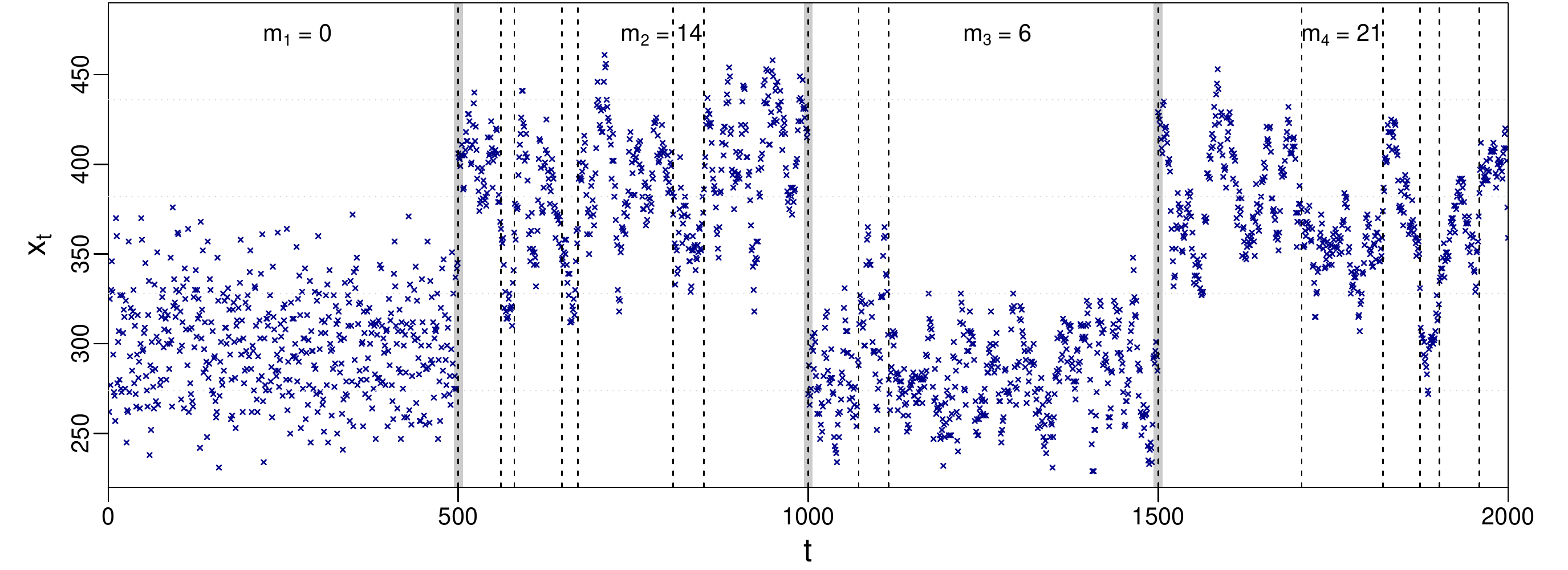}
\caption{Data generated from the moving-sum changepoint model for negative binomial data given in Example \ref{eg:negbin} with three changepoints ($\tau_1$, $\tau_2$ and $\tau_3$) indicated by thick grey lines. The black dashed lines indicate the positions of the MAP changepoints obtained by fitting the standard changepoint model (\ref{eq:classicconiid}) to the data.}
\label{fig:simplesimu}
\end{figure}

\subsection{Simulation from the model}
\label{sec:asimulation}
Figure \ref{fig:simplesimu} displays data generated from the moving-sum changepoint model, given a fixed sequence of changepoints, for negative binomial data given in Example \ref{eg:negbin} with $r = 200$, $\alpha = 20$ and $\beta = 10$. It is apparent that the changepoints correspond to changes in both the dependence structure and the mean of the data. In particular, we note that for larger values of $m_j$, the data tend to be smoother in the corresponding segment. For segments with $m_j>0$, it is reductive to judge the data to be exchangeable since there are clear temporal dynamics. 

The bottom panel of Figure \ref{fig:simplesimu} displays the positions of the MAP changepoints obtained by fitting the standard changepoint model for exchangeable data given in (\ref{eq:classicconiid}) to the simulated data using Metropolis-Hastings sampling of the changepoints as described in \citet{Denison}. 
Within segments where the data are not exchangeable, the standard model cannot capture the temporal dynamics and therefore the data are inferred to be more segmented than preferable.


\section{Conditional likelihood for the moving-sum changepoint model}
\label{sec:conditionallik}
Since changepoints split the data into independent segments, the joint posterior density of changepoints is tractable, up to a normalising constant, if the conditional likelihood of data within each segment can be computed. 
This section discusses the computation of the conditional likelihood of some data $x_1, \ldots, x_n$ observed within a single segment,  assuming the moving-sum model defined in Section \ref{sec:movingsumsegment} for some $m \geqslant 0$ and $\theta$.



\subsection{Relationship between the observed data and the latent variables} 
\label{sec:diffequation}
Before we propose one approach to obtain the conditional likelihood of the observed data within a generic segment, we give further insights on the relationship between the observed data and the latent variables. We show that, for the latent $m$-dependence framework (\ref{eq:movingsummodel}), there are $m$ free latent variables subject to some constraints, and then all further latent variables are implied by the observed data sequence. It will be notationally convenient to characterise the structure of the data within a segment with $n>1$ points using the sequence of finite backward differences, or backward jumps,  $\diffb x_2, \ldots, \diffb x_{n}$ with the backward difference operator $\diffb$ defined by
\begin{align}
\label{eq:diffbdef}
\diffb x_t = x_t - x_{t-1}.
\end{align}
Similarly, we define the sequence of forward differences $\difff x_2, \ldots, \difff x_{n}$ with the forward difference operator $\difff \, $ defined by 
\begin{align}
\label{eq:diffbdef2}
\difff \hspace{-0.4mm} x_t = x_t - x_{t+1} = - \diffb x_{t+1} .
\end{align}

The equation given in (\ref{eq:movingsummodel})  may be equivalently expressed as
\begin{align}
\label{eq:diffequation}
 y_t =  y_{t-\bar{m}} + \diffb x_t,
\end{align} 
for $t=2, \ldots, n$, and $y_1 = x_1 - (y_{-m+1} + \cdots + y_{0})$. Iterating the expression in (\ref{eq:diffequation}) shows that, given the initial $m$ latent random variables $y_{-m+1},  \ldots, y_{0}$, there is a one-to-one relationship between the finite differences of $x_{1:n}$ (\ref{eq:diffbdef}) and the remaining latent variables $y_{1:n}$. Let the first $m$ latent variables be $\gamma_{1:m}= (\gamma_1, \ldots, \gamma_m)$, with 
\begin{align}
\gamma_{r} = y_{-m+r}
\end{align}
for $r = 1, \ldots, m$. Explicitly, for all $t=1, \ldots, n$, letting $r$ be the remainder and $q$ the quotient of the Euclidean division of $t-1$ by $\bar{m}$ such that $t=q\bar{m}+r+1$, we have
\begin{align}
\label{eq:yxrela}
y_t = \left\lbrace \begin{array}{ll}
     \left( \sum_{i=0}^{q} \diffb x_{i\bar{m} +1 } \right) - \gamma^{\sbt},  & \text{$r=0$} \\
     \gamma_{r} - \sum_{i=0}^{q} \difff x_{i\bar{m} + r}, & \text{$r=1, \ldots, m$},
  \end{array} \right.
\end{align}
where $x_0 \equiv 0$ and 
\begin{align}
\label{eq:gammadot}
\gamma^{\sbt} = \sum_{r=1}^{m} \gamma_{r}.
\end{align}
The role played by $\gamma_{1:m}$ is akin to the role played by the unknown initial conditions of a stochastic difference equation.

The choice to condition on the first $m$ latent random variables is arbitrary; given any sequence of $m$ consecutive latent random variables, there is a one-to-one relationship between the other latent variables and the observed data. %
The following definition gives a transformation which may be used to obtain $\gamma_{1:m}$ from any $m$ consecutive latent random variables, such that it is sufficient to consider conditioning on the first $m$ latent variables in all subsequent discussion. The transformation is also useful in the later sections of the report, where, for example, we need to obtain the initial latent variables when some data are added to, or removed from,  an edge of a segment. 
\begin{defn}
\label{def:yinit}
Let $x_1, \ldots, x_n$ be data observed within a segment, assuming the model (\ref{eq:movingsummodel}) for some $m \geqslant 0$. Let $S[y_{t}, \ldots, y_{t+m-1} ] = (y_{t+1}, \ldots, y_{t+m} )$ denote the `shift' map, where $y_{t+m} = x_{t+m} - \sum_{i=0}^{m-1} y_{t+i} $, for all suitable $t$. Clearly, $S$ is iterable and invertible, and for all sequences of $m$ consecutive latent random variables $y_{(-m+1+u):u}$, with $ 0 \leqslant u \leqslant n$,  $S^{-u}[y_{(-m+1+u):u}] = y_{(-m+1):0} = \gamma_{1:m} $.
\end{defn}

\subsection{Conditional likelihood of the data within a segment}
\label{sub:condlik}

Given $m$ and $\gamma_{1:m}$, (\ref{eq:yxrela}) provides a one-to-one deterministic mapping between $x_{1:n}$ and $y_{1:n}$ with unit Jacobian. Hence, if we treat the sequence $\gamma_{1:m}$ as an additional unknown segment parameter, whose elements are independent and identically distributed with density $f_m(\cdot \, |\, \theta)$, then the conditional likelihood of the observed data within a segment is
\begin{align}
\label{eq:condilik}
L(x_{1:n} \, | \,\theta,  m, \gamma_{1:m}) &= L(y_{1:n} \, | \, \theta,  m,  \gamma_{1:m}) \\
& = \prod_{i=1}^{n} f_m(y_i | \theta).
\end{align}
Thus, using the notation introduced in Section \ref{sec:fullchangepointmodel}, but ignoring the subscripts corresponding to the indices of segments, the unknown segment parameters are $(\theta, m , \gamma_{1:m})$, with prior density $\pi(\theta) \pi(m) \pi(\gamma_{1:m} \, | \theta, m)$ where $\pi(\gamma_{1:m} \, | \theta, m) =  \prod_{r=1}^{m} f_m(\gamma_{r} \,  | \, \theta)$.

Recall it is assumed that the prior for $\theta$ is chosen to be conjugate for $f_m(\cdot \, |\, \theta)$ conditional on $m$. Consequently, the joint likelihood of the data and the initial latent variables conditional on $m$ can be derived by invoking Bayes' theorem,
\begin{align}
\label{eq:condilikmarg}
L(x_{1:n}, \gamma_{1:m} \, | \, m) = \int L(y_{1:n} \, | \, \theta, m,  \gamma_{1:m} )  \pi(\gamma_{1:m} \, | \, \theta, m) \pi(\theta) d \theta.
\end{align}

An expression for (\ref{eq:condilikmarg}) is given below for the three examplar segment models introduced in Section \ref{sec:movingsumsegment}, with $\mathcal{Y}_m \equiv \mathcal{Y}_m(x_{1:n})$ denoting the set of sequences $\gamma_{1:m}$ such that $y_t$ belongs to $\mathcal{F}$, the support of $f_m(\cdot \, |\, \theta)$, for all $t= 1, \ldots, n$, given $m$ and $x_{1:n}$. As stated in Remark \ref{remark:calY1}, it is not guaranteed that, for all $m$, there exist initial latent variables $\gamma_{1:m} \in \mathcal{Y}_m$ such that the conditional likelihood in (\ref{eq:condilikmarg}) is positive.

\begin{remark}[Set $\mathcal{Y}_m$]
\label{remark:calY1}
Note that in the case where $m=0$, meaning the sequence $x_{1:n}$ is assumed to be exchangeable, then $\gamma_{1:m}$ is the empty sequence, and the expression in (\ref{eq:condilik}) is always well defined. Now, if $m>0$, two cases need to be considered separately. If $\mathcal{F}$ is unbounded, for all $m$ and sequence $x_{1:n}$, the set $\mathcal{Y}_m$ is $\mathcal{F}^{m}$. However, if $\mathcal{F}$ is bounded then $\mathcal{Y}_m$ is a proper subset of $\mathcal{F}^{m}$ and is not necessarily non-empty for all $m>0$ and $x_{1:n}$. For example, 
if $\mathcal{F}$ is bounded below by 0, for any non-negative sequence $x_{1:n}$ with $x_2 > x_1 + x_3$, the set $\mathcal{Y}_m$ is empty for all $m>0$. 
\end{remark}

\addtocounter{exemp}{-3}

\begin{exemp}[Normal marginal distribution, continued]
\label{eg:normalgamma2}
Given parameters $( m , \gamma_{1:m})$ and a known hyperparameters $\lambda, \alpha, \beta >0$, it follows that
\begin{align*}
L(x_{1:n}, \gamma_{1:m} \,  | \,  m) &=  \left(  \frac{ \bar{m}  }{ 2 \pi } \right)^{( n+m)/2}   \left(  \frac{ \lambda }{ \lambda^\prime  } \right)^{1/2}  \frac{\beta ^{ \alpha } }{\Gamma( \alpha ) }   \frac{  \Gamma( \alpha^\prime ) }{  (\beta^{\prime })^{\alpha^\prime}  } ,
\end{align*}
where $\lambda^\prime = \frac{(n + m + \bar{m} \lambda )} { \bar{m} }$, $\alpha^\prime = \frac{ ( n+m) }{2} + \alpha$ and $\beta^\prime = \beta + \frac{ \bar{m} }{2} (\sum_{t=-m+1}^n y_t^2)   + \frac{ \lambda }{2} \mu_0^2 -  \frac{ (\lambda \mu_0 + \sum_{t=-m+1}^n y_t   )^2 }{  2 \lambda^\prime }$.
\end{exemp}

\begin{exemp}[Gamma marginal distribution, continued]
\label{eg:gamma2}
Given  parameters $( m , \gamma_{1:m})$ and known hyperparameters $\lambda, \alpha, \beta>0$, it follows that
\begin{align*}
L(x_{1:n}, \gamma_{1:m} \,  | \,  m) &= \frac{\Gamma( \alpha +(n+m)\lambda_{m}  )}{\Gamma(\alpha) \Gamma( \lambda_{m} )^{n+m} } \frac{\beta^{\alpha} y_{\sbt}^{(n+m)(\lambda_{m} -1 ) } }{(\beta + y^{\sbt} )^{\alpha +(n+m)\lambda_{m} }   } \frac{  }{  } \mathds{1}_{  \mathcal{Y}_m } (\gamma_{1:m}),
\end{align*}
where $\lambda_{m} = \lambda/\bar{m}$, $y_{\sbt}=\prod_{t=-m+1}^{n} y_t$ and $y^{\sbt}=\sum_{t=-m+1}^{n}y_t$. 
\end{exemp}

\begin{exemp}[Negative binomial marginal distribution, continued]
\label{eg:negbin2}
Given parameters $( m , \gamma_{1:m})$ and known hyperparameters $r, \alpha, \beta>0$, it follows that
\begin{align*}
L(x_{1:n}, \gamma_{1:m} \,  | \,  m) &= \frac{ \Gamma(\alpha + \beta) }{ \Gamma(\alpha) \Gamma(\beta)} \left( \prod_{t=-m+1}^n    \frac{\Gamma( y_t + r_m )}{ \Gamma( y_t + 1) \Gamma(r_m) } \right) \frac{ \Gamma( y^{\sbt} + \alpha ) \Gamma( (n+m)r_m + \beta  )  }{ \Gamma( y^{\sbt} + \alpha + (n+m)r_m + \beta ) }\mathds{1}_{  \mathcal{Y}_m } (\gamma_{1:m}),
\end{align*}
where $r_{m} = r/\bar{m}$ and $y^{\sbt}=\sum_{t=-m+1}^{n}y_t$.
\end{exemp}

The segment parameters $(m, \gamma_{1:m})$ cannot be marginalised, and consequently to sample from the posterior distribution of the changepoints, we need to sample $(m, \gamma_{1:m})$ for each segment. Different challenges arise when attempting to do so: the dimension of the segment parameter space is unknown; as first hinted in Remark \ref{remark:calY1}, the parameter space depends on the observed data when the support of $f_m(\cdot \, |\, \theta)$ is bounded. Hence, in order to develop a sampling strategy which is computationally realistic, it is necessary to characterise the parameter space we are seeking to navigate.

\section{Analysis of the latent parameter space}
\label{sec:parameterspace}
In Section \ref{sub:condlik} we defined $\mathcal{Y}_{m}$ to be set of sequences $\gamma_{1:m}$ such that the joint conditional probability density of $\gamma_{1:m}$ and $x_{1:n}$ is non-zero within a generic segment. We now define 
\begin{align}
\mathcal{M} \equiv \mathcal{M}(x_{1:n}) = \{ m \in \mathbb{N}_{0} \, | \, \mathcal{Y}_{m} \neq \emptyset \}
\end{align}
to be the set of $m \geqslant 0$ for which $\mathcal{Y}_{m}$ is non-empty. In other words, given some observed data $x_1, \ldots, x_n$ within a generic segment, $\mathcal{M}$ and $\mathcal{Y}_{m}$ provide the values of $m$ and $\gamma_{1:m}$ for which the segment model (\ref{eq:movingsummodel}) is valid. 

As stated in Remark \ref{remark:calY1}, it is always possible to assume the sequence $x_{1:n}$ to be exchangeable, and hence $0\in \mathcal{M}$. If $\mathcal{F}$, the support of $f_m(\cdot \, |\, \theta)$,  is unbounded, such as in Example \ref{eg:normalgamma2}, then it is immediate that $\mathcal{M} = \mathbb{N}_{0}$, and $\mathcal{Y}_{m}$ is $\mathcal{F}^{m}$ for all $m>0$. However, if $\mathcal{F}$ is bounded, such as in Example \ref{eg:gammagamma} and Example \ref{eg:negbin2}, then both $\mathcal{M}$ and $\mathcal{Y}_{m}$ depend on the observed data $x_{1:n}$, and $\mathcal{M}$ and $\mathcal{Y}_{m}$ are proper subsets of $\mathbb{N}_{0}$ and $\mathcal{F}^{m}$, respectively. Note that infinitely divisible distributions with support bounded from below and from above have zero variance \citep{Steutel}, and therefore we only consider the case where $\mathcal{F}$ is bounded below but not above, without loss of generality. 

In this section, we explicitly state the relationship between the observed data $x_{1:n}$ and the sets $\mathcal{M}$ and $\mathcal{Y}_{m}$ for $m\in \mathcal{M}$ when $\mathcal{F}$ is bounded below, paving the way to designing a sampling strategy for their posterior distributions, which exploits the structure of the constrained parameter space.

\label{sec:parameterspace}

\subsection{Characterisation of the parameter space in terms of the observed data}
\label{sec:characparameterspace}

Suppose that $\mathcal{F}$ is unbounded above but bounded below by a constant, which can be set to $0$ without loss of generality. It follows that, given a sequence $x_{1:n}$ and some $m>0$, the set $\mathcal{Y}_{m}$ consists of those $\gamma_{1:m} \in \mathcal{F}^{m}$ such that $y_{t} \geqslant 0$ for all $t=1, \ldots, n$. 
According to (\ref{eq:yxrela}) and (\ref{eq:gammadot}), defining
\begin{align}
\label{eq:yinitbounds}
 \begin{array}{l} 
     U^{m} \equiv U^{m}(x_{1:n}) = \min \{ \diffb x_1 , \ldots, \sum_{q = 0 }^{\kappa_{\bar{m}}}  \diffb x_{q\bar{m}+1}  \}, \vspace{1.5mm} \\
   L_r^{m} \equiv  L_{r}^{m}(x_{1:n}) = \max \{ 0, \difff x_{r} , \ldots, \sum_{q = 0 }^{\kappa_{r}}  \difff x_{q\bar{m}+r}  \},
\end{array} 
\end{align}
where $x_0 \equiv 0$ and $\kappa_r = \kappa_r(n)$ is the largest $q \in \mathbb{N}_{0}$ such that $q\bar{m}+r+1 \leqslant n$ for all $r$, then
\begin{align}
\label{eq:mathcaldef}
\mathcal{Y}_{m} = \mathds{Y}[U^{m}, L^{m}_{1:m}] = \{ \gamma_{1:m} \, | \, U^{m} \geqslant \gamma^{\sbt} \text{ and } \gamma_r \geqslant L_r^{m} \text{ for all }  r=1, \ldots, m \}.
\end{align}
Although it might seem artificial to characterise $\mathcal{Y}_{m}$ is terms of $\mathds{Y}$ at this stage, such a representation will be useful both in the remainder of this section and in Section \ref{sec:rjmcmc}.
Moreover, for $m\geqslant 0$,  $m\in \mathcal{M}$ and $\mathcal{Y}_{m}$ is non-empty if and only if 
\begin{align}
\label{eq:mathcalMdef}
D^m \equiv D^{m}(x_{1:n}) = U^{m} - \sum_r L_{r}^{m} \geqslant 0.
\end{align}

Example \ref{eg:bounds1} gives expanded expressions of the bounds of $\mathcal{Y}_{m}$ for a small sequence of observed data. 

\begin{exemp}[Bounds of $\mathcal{Y}_{m}$]
\label{eg:bounds1}
Let $x_1, \ldots, x_{7}$ be count data. If $m=1$ ($\bar{m}=2$), then 
\begin{align*}
\left\lbrace \begin{array}{l}
      U^{1} = \min\{ x_1, x_1 + (x_3 - x_2), x_1 + (x_3 - x_2) + (x_5 - x_4), x_1 + (x_3 - x_2) + (x_5 - x_4) + (x_7-x_6) \}  \\
     L_1^{1} = \max\{ 0, (x_1 - x_2 ), (x_1 - x_2 ) + (x_3 - x_4 ), (x_1 - x_2 ) + (x_3 - x_4 ) + (x_5 - x_6 ) \},
 \end{array} \right.
\end{align*}
whilst if $m=3$  ($\bar{m}=4$), then 
\begin{align*}
\left\lbrace \begin{array}{l}
\, U^{3} \, = \min\{ x_1, x_1 + (x_5 - x_4)  \} \vspace{1mm} \\
\vspace{1mm} L_1^{3} = \max\{ 0, (x_1- x_2), (x_1- x_2) + (x_5 -x_6)  \}\\
 \vspace{1mm} L_2^{3} = \max\{ 0, (x_2- x_3),  (x_2- x_3) + (x_6-x_7) \} \\
\vspace{1mm} L_3^{3} = \max\{ 0, (x_3- x_4) \}.
 \end{array} \right.
\end{align*}
\end{exemp}

\definecolor{myblue}{rgb}{0.0, 0.28, 0.67}

\begin{figure}[!t]
\centering
{
\small

\begin{tikzpicture}
\draw[->, line width=0.2mm,color=gray!90](0,-1) -- (0,6);
\draw[line width=0.1mm,color=gray](0.05,0) -- (-0.05,0) node[anchor= east]{$0$};
\draw[line width = 0.4mm, color=myblue, draw opacity = 0.5] (0,1.5)--(0,4.5);
\draw[line width=0.1mm,color=black](0.05,1.5) -- (-0.05,1.5) node[anchor= east]{$L^1_1$};
\draw[line width=0.1mm,color=black](0.05,4.5) -- (-0.05,4.5) node[anchor= east]{$U^1$};
\draw (0, 2.7) node[circle, fill=black, inner sep = 0.4mm]{};
\draw (0, 2.7) node[left]{$\gamma_1$};

\draw[->,line width=0.3mm,myblue] (0.6, 5.2) node[above]{$\mathcal{Y}_1$} to[out=-90,in=60] (0, 3.7) ;

\draw (0,-2) node[]{$m=1$};
\end{tikzpicture}
\hspace{-1mm}
\begin{tikzpicture}
\draw[->, line width=0.2mm,color=gray!90](0,-1) -- (0,6);
\draw[->, line width=0.2mm,color=gray!90](-0.8,0) -- (4.2,0);
\draw (0,0) node[anchor=north east, color=gray!90]{$0$};

\draw[dashed, line width=0.05mm] (0.5,0) -- (0.5,5); 
\draw[dashed, line width=0.05mm] (0,0.8) -- (3.7,0.8); 
\draw[dashed, line width=0.05mm] (0,3.8) -- (3.8,0); 
\draw[dashed, line width=0.05mm] (0,3.3) -- (0.5,3.3); 
\draw[dashed, line width=0.05mm] (3,0) -- (3,0.8); 

\draw[line width=0.1mm,color=black](0.05,3.8) -- (-0.05,3.8) node[anchor= east]{$U^2$};
\draw[line width=0.1mm,color=black](3.8, 0.05) -- (3.8, -0.05) node[anchor= south, inner sep=1.1mm, xshift=2.5mm]{$U^2$};

\draw[line width=0.1mm,color=black](0.05,3.3) -- (-0.05,3.3) node[anchor= east]{$U^2-L^2_2$};
\draw[line width=0.1mm,color=black](0.05,0.8) -- (-0.05,0.8) node[anchor= east]{$L^2_1$};

\draw[line width=0.1mm,color=black](3, 0.05) -- (3,-0.05) node[anchor= north]{$U^2-L^2_1$};
\draw[line width=0.1mm,color=black](0.5, 0.05) -- (0.5, -0.05) node[anchor= north]{$L^2_2$};

\draw[line width = 0.15mm, color=black, fill=myblue, fill opacity=0.3] (0.5, 0.8)-- (3,0.8) -- (0.5, 3.3) --cycle;

\draw (1.3, 1.9) node[circle, fill=black, inner sep = 0.4mm]{};
\draw[dotted, line width=0.1mm] (1.3, 1.9) -- (0, 1.9) node[left]{$\gamma_1$};
\draw[dotted, line width=0.1mm] (1.3, 1.9) -- (1.3, 0) node[below]{$\gamma_2$};

\draw[->,line width=0.3mm,myblue] (2.2, 4.9) node[above]{$\mathcal{Y}_2 = Y[U^2, L^2_{1:2}]$} to[out=-90,in=60] (0.9, 2.4) ;

\draw (2,-2) node[]{$m=2$};
\end{tikzpicture}
\hspace{-10mm}
\begin{tikzpicture}

\draw[->, line width=0.2mm,color=gray!90](0,0) -- (0,6);
\draw[->, line width=0.2mm,color=gray!90](0, 0) -- (5.7,-1.5);
\draw[->, line width=0.2mm,color=gray!90](0, 0) -- (5.7,1.5);
\draw (0,0) node[anchor=east, color=gray!90]{$0$};

\draw[dashed, line width=0.05mm](0, 1) -- (5.1,-0.28);
\draw[dashed, line width=0.05mm](0, 1) -- (5.1,2.4);
\draw[line width=0.1mm,color=black](0.05, 1) -- (-0.05, 1) node[anchor= east]{$L^3_1$};

\draw[dashed, line width=0.05mm](0.8, 0.8) -- (5.1,1.97);
\draw[dashed, line width=0.05mm](0.8, 0.73) -- (0.8,-0.30);
\draw[line width=0.1mm,color=black](0.8,-0.18) -- (0.8,-0.3) node[anchor= north]{$L^3_2$};

\draw[dashed, line width=0.05mm](0.51, 1.14) -- (5.0,0.09);
\draw[dashed, line width=0.05mm](0.51, 1.14) -- (0.51,0.08) node[anchor= south east, inner sep=0.19mm]{$L^3_3$};

\draw[dashed, line width=0.05mm] (3.28,0.18) -- (3.79,0.37);
\draw[dashed, line width=0.05mm] (3.28,0.18) -- (3.28,-0.9);
\draw[line width=0.1mm,color=black](3.28,-0.85) -- (3.28,-0.95) node[anchor= south west, inner sep=0.5mm]{$U^3-(L^3_1+L^3_3)$};

\draw[dashed, line width=0.05mm] (3.79,0.37) -- (3.95,1.65);
\draw[ dashed, line width=0.05mm] (3.95,1.65) -- (3.15,1.865);

\draw[ dashed, line width=0.05mm] (1.35,0.95) -- (1.35,3.6);

\draw[line width = 0.15mm, color=black, fill=myblue, fill opacity=0.3] (1.35,0.95)-- (3.79,0.37) -- (3.95,1.65) --cycle;

\draw[line width = 0.15mm, color=black, fill=myblue, fill opacity=0.3] (1.35,0.95)-- (1.35,3.6) -- (3.95,1.65) --cycle;

\draw[line width = 0.15mm, color=black, fill=myblue, fill opacity=0.3] (1.35,0.95)-- (1.35,3.6) -- (3.79,0.37) --cycle;

\draw[dashed, line width=0.01mm, color = gray!30](0, 1+3.6-0.95) -- (5.1,3.6-0.95+-0.28);
\draw[dashed, line width=0.01mm, color = gray!30](0,1+3.6-0.95) -- (5.1, 3.6-0.95+2.4);

\draw[dashed, line width=0.05mm] (0,1+3.6-0.95) -- (0.51 + 1.9,1.14+3.6-0.95 + 0.52 );
\draw[dashed, line width=0.05mm] (0,1+3.6-0.95) -- (0.8 + 0.35, 0.8+3.6-0.95 -0.08 );

\draw[dashed, line width=0.05mm] (0.8,0.8+3.6-0.95) -- (1.35,3.6);

\draw[dashed, line width=0.05mm] (0.51,1.14+3.6-0.95) -- (1.35,3.6);

\draw[line width=0.1mm,color=black](0.1,1+3.6-0.95) -- (-0.1,1+3.6-0.95) node[anchor= north east, inner sep=0.5mm]{$U^3-(L^3_2+L^3_3)$};

\draw[->,line width=0.3mm,myblue] (3.1, 4.9) node[above]{$\mathcal{Y}_3 = Y[U^3, L^3_{1:3}]$} to[out=-90,in=60] (2.5, 2.4) ;

\draw (2.7,-2) node[]{$m=3$};

\end{tikzpicture}

\caption{Cartoons of the set $\mathcal{Y}_m$ for $m=1,2,3$ with $\mathcal{F}$ assumed to be continuous.}
\label{fig:calYn}
}

\end{figure}
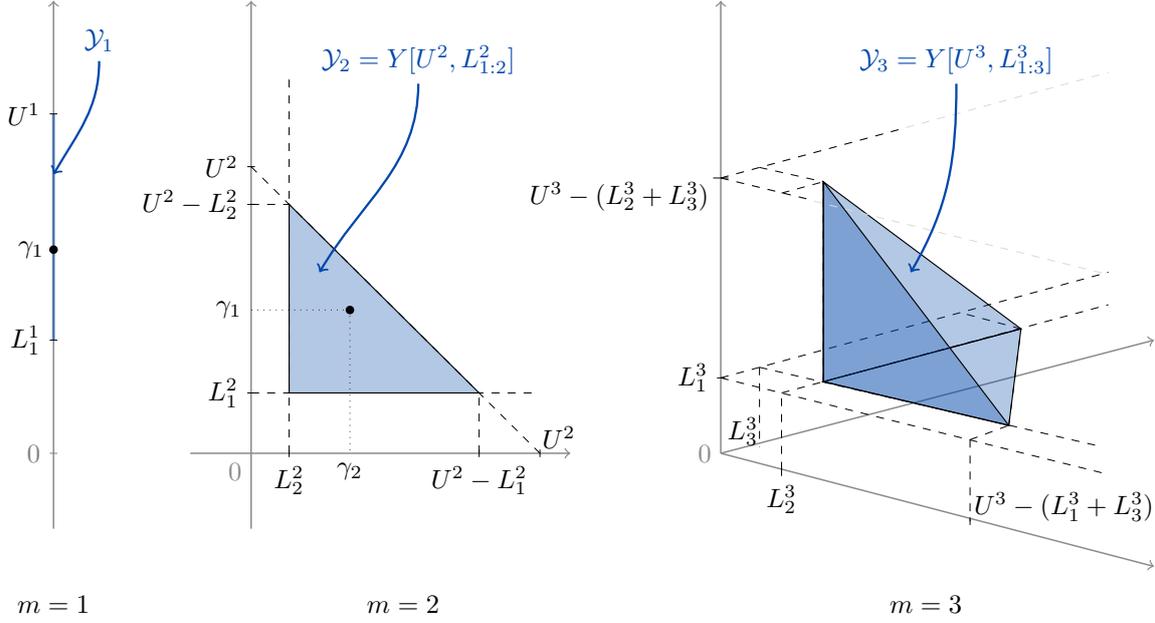

Figure \ref{fig:calYn} displays cartoon representations of $\mathcal{Y}_{m}$ for $m=1,2$ and $3$. For all $m>0$, if non-empty, $\mathcal{Y}_{m}$ is a convex polyhedron in $\mathcal{F}^{m}$, whose vertices are determined by the bounds $U^{m}$ and $L^{m}_{1:m}$, and whose size only depends on $m$ and $D^{m}$ for all $m$.

Given a sequence of observed data $x_{1:n}$ within a segment, it was shown in (\ref{eq:mathcaldef}) that the sets $\mathcal{M}$ and $\mathcal{Y}_{m}$ may be expressed in terms of the first observation $\diffb x_1 =x_1 - x_0 = x_1$ and all of the subsequent the finite differences of the data. Essentially, for all $m$, the larger the finite differences, or \textit{jumps}, are relative to $x_1$, the smaller $\mathcal{Y}_m$ becomes unless the jumps happen to exhibit negative $\bar{m}$-lagged autocorrelations. In particular, if the observed data $x_{1:n}$ consist of a succession of gradual drifts, whose directions flip every $\bar{m}$ time points, then  $\mathcal{Y}_m$ is large with respect to $x_1$. Furthermore, in order to stress that the size of the jumps must be considered relative to $x_1$, we note that for all sequences $x_{1:n}$ such that $\mathcal{Y}_m$ is non empty for some $m$, and for all constant $\mu$, $D^{m}( x_{1:n} + \mu ) = D^{m}( x_{1:n} )+ \mu$, where $x_{1:n} + \mu$ denotes the sequence $(x_{t} + \mu)_{t=1}^{n}$.

\subsection{Adding and removing data}
\label{sec:addremdata}

To develop a sampling strategy for the latent variables, it is important to determine how $\mathcal{M}$ and $\mathcal{Y}_m$, $m\in \mathcal{M}$, may change when data are added to, or removed from, an edge of a segment. Proposition \ref{prop:addremdatawithproof} guarantees that, for all sequences of data, if there exist segment parameters $\gamma_{1:m}$ for some $m$, then we can obtain valid  parameters of the same dimension for all sequences of data which could be obtained by removing data from the beginning or the end of the original sequence.

\begin{propr}
\label{prop:addremdatawithproof}
Let $x_{1:n}$ be observed data and $ 1 \leqslant s \leqslant t \leqslant n$. Then $\mathcal{M}(x_{1:n}) \subseteq \mathcal{M}(x_{s:t})$, and $S^{s-1}[\mathcal{Y}_{m}(x_{1:n})]\subseteq \mathcal{Y}_{m}(x_{s:t})$ for all $m\in \mathcal{M}(x_{1:n})$, where the shift map $S$ is defined in Definition \ref{def:yinit}. In particular, if $s=1$ we have $\mathcal{Y}_{m}(x_{1:n})\subseteq \mathcal{Y}_{m}(x_{1:t})$. 
\begin{proof}
Suppose $\gamma_{1:m} = y_{(-m+1):0} \in \mathcal{Y}_{m}(x_{1:n})$ for some $m \in \mathcal{M}(x_{1:n})$. Let $y_{1:n}$ denote the latent variables obtained from $\gamma_{1:m}$ and $x_{1:n}$ via (\ref{eq:yxrela}). By definition of the shift map, the latent variables obtained from $S^{s-1}[y_{(-m+1):0}] = y_{(-m+s):(s-1)}$ and $x_{s:t}$ via (\ref{eq:yxrela}) are equal to $y_{s:t}$. Hence $S^{s-1}[y_{(-m+1):0}]  \in \mathcal{Y}_{m}(x_{s:t})$, and therefore $m \in \mathcal{M}(x_{s:t})$.
\end{proof}
\end{propr}
However, if some data are added to the beginning or the end of a sequence of data for which we currently have  valid parameters $\gamma_{1:m}$ for some $m$, then it is not guaranteed that $m \in \mathcal{M}$ for the extended sequence of data. For example, it follows directly from (\ref{eq:yinitbounds}) that, as more data are added to the end of a segment, for all $m$, $U^{m}$ may only decrease and $L^{r}_m$ may only increase for all $r$, such that $\mathcal{Y}_{m} = \mathds{Y}[U^{m}, L^{m}_{1:m}]$ may only shrink.

\subsection{Different ranges of dependence}
\label{sec:transdimenjump}
Given a segment of data, there is a non-negligible computational cost in verifying via (\ref{eq:mathcalMdef}) that some integer $m$ belongs to the corresponding set $\mathcal{M}$. Hence, it is instructive to determine whether knowing that some $m$ belongs to $\mathcal{M}$ can inform whether some $m^\prime \neq m$ is also an element of $\mathcal{M}$.  

\begin{figure}[t!]
\centering
\begin{tikzpicture}
\draw[-, line width=0.2mm] (90: 0) -- (90: 4);
\draw[-, line width=0.2mm] (60: 0) -- (60: 4);
\draw[-, line width=0.2mm] (30: 0) -- (30: 4);
\draw[-, line width=0.2mm] (0: 0) -- (0: 4);
\draw[-, line width=0.2mm] (-30: 0) -- (-30: 4);

\draw (90: 2.7) node[label=left:$\color{red} \gamma_1$]{$ \color{red} \times$};
\draw (60: 1.8) node[label=left:$\color{blue} \gamma_2$]{$ \color{blue}\times$};
\draw (30: 2.7) node[label=left:$\gamma_3$]{$\times$};
\draw (0: 2.9) node[label=above:$\color{red} \gamma_4$]{$ \color{red} \times$};
\draw (-30: 2.4) node[label=above:$\color{blue} \gamma_5$]{$ \color{blue}\times$};

\draw[dotted, line width=0.2mm, color=gray!80](90: 4) -- (60: 4);
\draw[dotted, line width=0.2mm, color=gray!80](90: 4) -- (30: 4);
\draw[dotted, line width=0.2mm, color=gray!80](90: 4) -- (0: 4);
\draw[dotted, line width=0.2mm, color=gray!80](90: 4) -- (-30: 4);

\draw[dotted, line width=0.2mm, color=gray!80](60: 4) -- (30: 4);
\draw[dotted, line width=0.2mm, color=gray!40](60: 4) -- (0: 4);
\draw[dotted, line width=0.2mm, color=gray!40](60: 4) -- (-30: 4);

\draw[dotted, line width=0.2mm, color=gray!80](30: 4) -- (0: 4);
\draw[dotted, line width=0.2mm, color=gray!40](30: 4) -- (-30: 4);

\draw[dotted, line width=0.2mm, color=gray!80](0: 4) -- (-30: 4);

\draw[-, line width=0.2mm](8,0-1) -- (12,0-1);
\draw[-, line width=0.2mm](8,0-1) -- (8,4-1);

\draw[dotted, line width=0.2mm, color=gray!90](8,4-1) -- (12,0-1);

\draw (9,0-1) node[]{$\color{blue}\times$};
\draw (9,-1.5) node[]{$\color{blue}\gamma_2^{\prime} = \gamma_2 + \gamma_5$};
\draw (8,0.7) node[]{$\color{red}\times$};
\draw (6.7,0.7) node[]{$\color{red}\gamma_1^{\prime}= \gamma_1+\gamma_4$};

\draw (2, 4.5) node[]{$\gamma_{1:5} \in \mathcal{Y}_5$};
\draw (10, 4.5) node[]{$\gamma_{1:2}^{\prime} = J_2(\gamma_{1:5}) \in \mathcal{Y}_2$};

\end{tikzpicture}
\caption{Cartoon representation of the mapping of $\gamma_{1:m} \in \mathcal{Y}_m$ to $\gamma_{1:{m^\prime}}^{\prime} = J_{m^\prime}(\gamma_{1:m}) \in \mathcal{Y}_{m^\prime}$ with $m=5$ and $m^\prime=2$. }
\label{fig:transJ}
\end{figure}
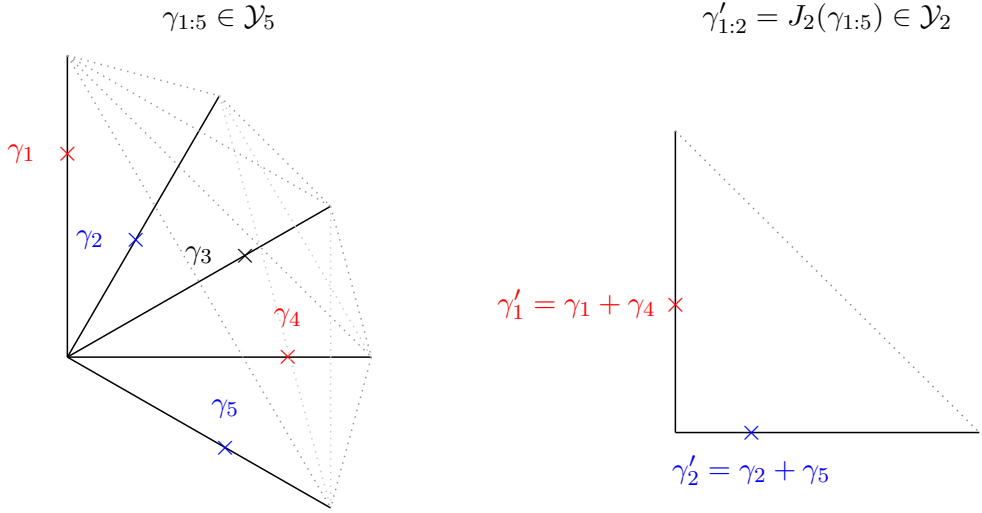

Suppose $m \in \mathcal{M}$ and $0 \leqslant m^\prime \leqslant m$ such that $\bar{m}^{\prime}$ divides $\bar{m}$ (written $\bar{m}^{\prime} \, | \,  \bar{m}$), meaning $\bar{m} = \ell \bar{d}$ for some $\ell>0$. Then define a mapping 
\begin{align}
J_{m^\prime }(\gamma_{1:m}) = \left( \sum_{j=0}^{\ell-1} \gamma_{j\bar{m}^\prime+1}, \ldots, \sum_{j=0}^{\ell-1} \gamma_{j\bar{m}^\prime+m^{\prime}}  \right)
\end{align}
for aggregating the latent variables. Figure \ref{fig:transJ} displays a cartoon representation of the mapping.  This mapping is required for the following proposition. 

\vspace{5mm}

\begin{propr}
\label{prop:mdjump}
For all $m \in \mathcal{M}$, $\gamma_{1:m} \in \mathcal{Y}_m$ and $m^\prime < m$ such that $\bar{m}^\prime \,  | \,  \bar{m}$,
\begin{itemize}
\item[(i)] $m^{\prime} \in \mathcal{M}$ and $J_{m^{\prime}}(\gamma_{1:m}) \in \mathcal{Y}_{m^{\prime}}$.
\item[(ii)] $J_{m^{\prime}}(\mathcal{Y}_m )  = \mathds{Y}[\tilde{U} , J_{m^{\prime}}(L^{m}_{1:m} )] \subset \mathds{Y}[U^{{ m^{\prime}}}, L^{ m^{\prime} }_{1:m^{\prime}}] = \mathcal{Y}_{m^{\prime}}$, where $\tilde{U} = U^{m}- \sum_{j=1}^{\ell-1} L^{m}_{j\bar{m}^\prime}$. 
\end{itemize}
\begin{proof}
See Appendix \ref{appen:proofs}.
\end{proof}
\end{propr}

To provide some intuition for Proposition \ref{prop:mdjump}, it is helpful to consider the bounds given in Example \ref{eg:bounds1}, noting that $3+1 = 2(1+1)$ and that the finite differences appearing in $L^{3}_1$ and $L^{3}_3$ for $m=3$ coincide with the finite differences in $L^{1}_1$ for $m=1$. 

Proposition \ref{prop:mdjump} says that for all $m \in \mathcal{M}$ it follows immediately that $\mathcal{D}(m) \subset \mathcal{M}$, where $\mathcal{D}(m)$ consists of the integers $m^\prime$ such that $\bar{m}^\prime$ divides $\bar{m}$.

\subsection{Asymptotic properties of the parameter space}
\label{sec:asymptoparam}
In Section \ref{sec:addremdata} it was shown that the parameter space may only shrink as more data are observed within a segment. Proposition \ref{prop:almostsureconvergence} sheds further light upon the asymptotic properties of $\mathcal{M}$ and $\mathcal{Y}_m$ for all $m$.

\begin{propr}
\label{prop:almostsureconvergence}
For $m \in \mathbb{N}_{0}$, suppose a sequence of latent variables $y_{-m+1}, \ldots, y_{n}$, and a sequence of observed data $x_{1}, \ldots, x_n$ are generated from model (\ref{eq:movingsummodel}), assuming some density $f_m(\cdot \, |\, \theta)$ with support $\mathbb{N}_{0}$ or $\mathbb{R}^{+}$. As $n \rightarrow \infty$, for all $m^\prime>0$,
\begin{align*}
U^{m^\prime}(x_{1:n}) & \overset{\text{a.s.}}{\xrightarrow{\hspace*{0.5cm}}}  \left\lbrace \begin{array}{cl}
     \sum_{r=1}^{m^\prime} \gamma^{\prime}_r  & \text{if } \,  m^\prime \in \mathcal{D}(m) \\
     - \infty & \text{otherwise}
 \end{array} \right.   \\
  \quad L^{m^\prime}_r(x_{1:n}) & \overset{\text{a.s.}}{\xrightarrow{\hspace*{0.5cm}}}  \left\lbrace \begin{array}{cl}
      \gamma^{\prime}_r  & \text{if } \,  m^\prime \in \mathcal{D}(m) \\
      \infty & \text{otherwise}
 \end{array} \right.
\end{align*}
for all $r=1, \ldots, m^\prime$, with $\gamma^{\prime}_{1:m^\prime} = J_{m^\prime}(\gamma_{1:m})$ and $\mathcal{D}(m) = \{ m^\prime \in \mathbb{N}_{0},  \, \,    \bar{m}^\prime \,  | \,  \bar{m}  \}$, such that, almost surely, $\mathcal{M}(x_{1:n})$ converges to $\mathcal{D}(m)$ and, for all $m^\prime \in \mathcal{D}(m)$, $\mathcal{Y}_{m^\prime}(x_{1:n}) = \mathds{Y}[U^{m^\prime}, L^{m^\prime}_{1:m^\prime}] $ converges to $\{ \gamma^{\prime}_{1:m^\prime} \}$.
\begin{proof}
See Appendix \ref{sec:apendproofpropconvergence}.
\end{proof}
\end{propr}

\begin{figure}[t!]
\centering
\includegraphics[height=0.7\textwidth]{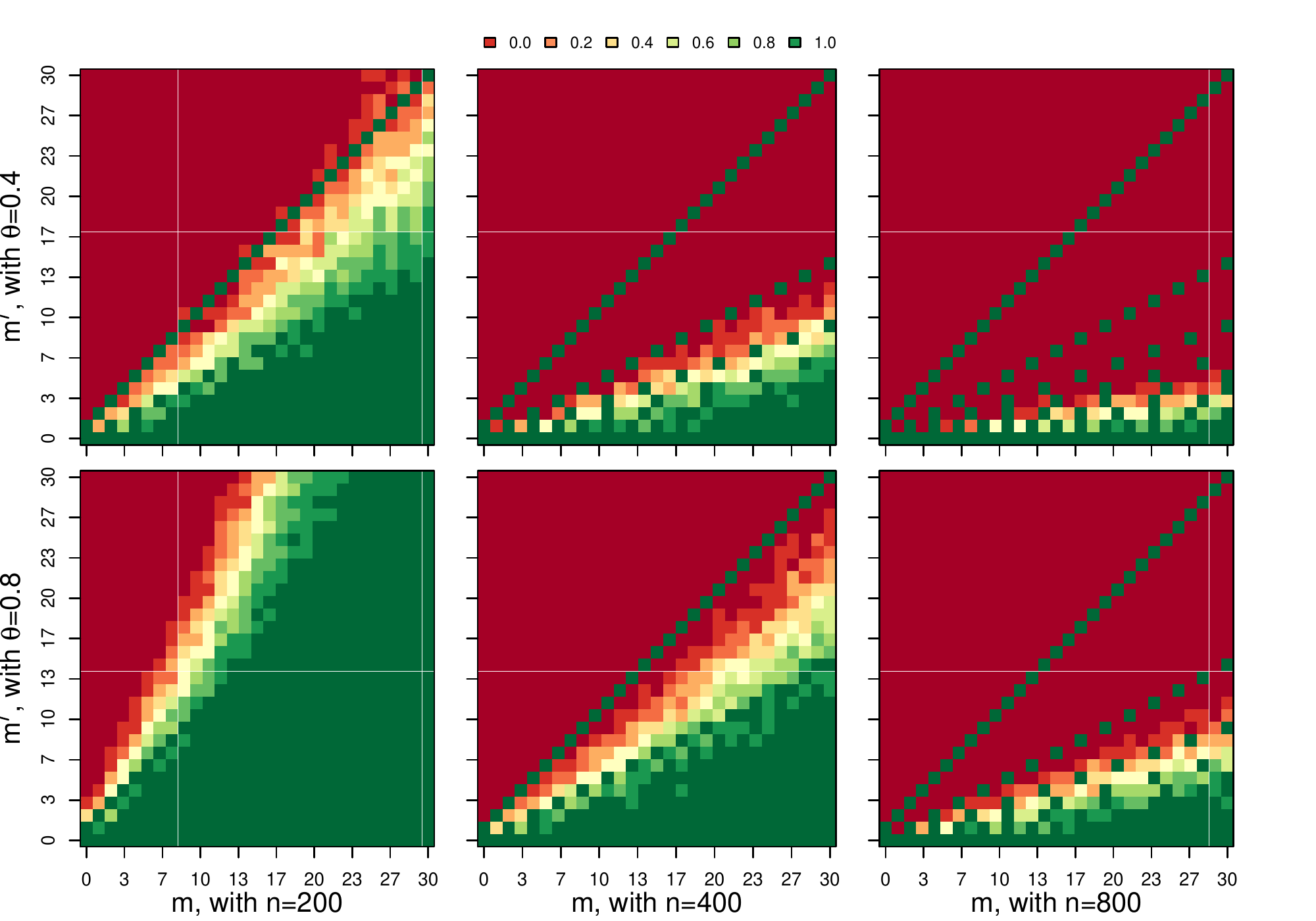}
\caption{The colours indicate the proportion of simulations for which $m^\prime \in \mathcal{M}$, denoted $Q( m^\prime \in \mathcal{M} \, | \, n, m, \theta )$, based on 50 simulations from the moving-sum segment model for negative binomial data given in Example \ref{eg:negbin} for $m=0, \ldots, 30$, $n=200, 400, 800$ and $\theta = 0.4, 0.8$.}
\label{fig:pspacesimu}
\end{figure}

Proposition \ref{prop:almostsureconvergence} tells us that if some data are generated from model (\ref{eq:movingsummodel}) for some $m>0$ and some initial latent variables $\gamma_{1:m}$, then almost surely, as the number of observations tends to infinity, $\mathcal{M}$ converges to $\mathcal{D}(m)$ and $\mathcal{Y}_{m^\prime}$ converges to a set containing a unique sequence, namely the transformation of $\gamma_{1:m}$ by $J_{m^\prime}$, for all  $m^\prime \in \mathcal{D}(m)$. We note that the result shows that both the structure in $\mathcal{M}$ and the transformation identified in Proposition \ref{prop:mdjump} are fundamental, and increasingly important as the length of a changepoint segment increases.

Within the changepoint detection framework, it is appealing that, the more data are observed within a segment, the more information the parameter space gives us about the nature of the dependence within the segment, so that a changepoint may be forced immediately upon observing some data generated from a different dependency structure. 

Recall that in Section \ref{sec:diffequation} it was argued that the segment parameters $\gamma_{1:m}$ may be considered as the unknown initial conditions of a process governed by a difference equation. From this point of view, Proposition \ref{prop:almostsureconvergence} says that, asymptotically, the uncertainty on the initial conditions vanishes.

An experiment was performed to illustrate Proposition \ref{prop:almostsureconvergence}. For six different parameter configurations corresponding to different fixed values of $n$ and $\theta$, and for all $m$ in $\{ 0, \ldots, 30 \} $, we performed 50 simulations from the moving-sum segment model for negative binomial data given in Example \ref{eg:negbin} with $r=300$. For all $m^\prime\in \{ 0, \ldots, 30 \}$ we computed the proportion of simulations for which $m^\prime \in \mathcal{M}$, denoted $Q( m^\prime \in \mathcal{M} \, | \, n, m, \theta )$. Figure \ref{fig:pspacesimu} displays the results of the experiment. As expected from Proposition \ref{prop:almostsureconvergence}, for all parameter choices it is apparent that $m^\prime \in \mathcal{M}$ with estimated probability $1$ for all $m^\prime \in \mathcal{D}(m)$, and that $Q( m^\prime \in \mathcal{M} \, | \, n, m, \theta )$ tends to decrease as $n$ increases for all $m^\prime \notin \mathcal{D}(m)$. Moreover, three other trends are worth mentioning. First, $Q( m^\prime \in \mathcal{M} \, | \, n, m, \theta )$ tends to increase as $m$ increases for all $m^\prime \notin \mathcal{D}(m)$ given $n$ and $\theta$. Second, it tends to be more likely for $m^\prime_1$ to be in $\mathcal{M}$ than for $m^\prime_2$ to be in $\mathcal{M}$ for all $m^\prime_1, m^\prime_2 \notin \mathcal{D}(m)$ such that $m^\prime_1 < m^\prime_2$. Third, one may observe that $Q( m^\prime \in \mathcal{M} \, | \, n, m, \theta )$ tends to increase as $\theta$ increases for all $m^\prime \notin \mathcal{D}(m)$ given all $m$ and $n$.

\section{Markov chain Monte Carlo changepoint inference}
\label{sec:rjmcmc}

For the Bayesian changepoint model given in Section \ref{sec:fullchangepointmodel}, the reversible jump MCMC algorithm \citep{Green1995}, which is a Metropolis-Hastings algorithm suitable for target distributions of varying dimension, may be used to sample from the posterior distribution of the positions of an unknown number of changepoints. Four types of moves are considered to explore the support of the target distribution: \textit{shift} of a randomly selected changepoint; change of a randomly chosen segment parameter; \textit{birth} of a new changepoint chosen uniformly over the time period; and \textit{death} of a randomly selected changepoint.

In this section, within the framework given in \citet{Green1995}, we propose a strategy to sample from the posterior distribution of  changepoints when the moving-sum model defined in Section \ref{sec:movingsummodelfull} is assumed for each segment. To address the challenges that the dimensions of the segment parameters are unknown and that the segment parameter space depends on the observed data within each segment, we exploit the analysis of the segment parameter space in Section \ref{sec:parameterspace}. 

\subsection{Description of the sampler}
\label{sec:sampler}
Suppose that the latest particle of the sample chain consists of $k$ changepoints, whose positions are $\tau_{1:k}$, and $(k+1)$ segment parameters. For segment $j$, recall from Section \ref{sub:condlik} that we assume the latent variable density parameter $\theta_{j}$ may be marginalised, such that the segment parameters consist of the order of dependence $m_{j}$ and the initial latent variables $\gamma_{ j,:} \equiv  \gamma_{ j,1:m_j} = (\gamma_{j,1}, \ldots, \gamma_{j,m_j})$.  
To explore the support of the target distribution, we propose the next element of the chain via one of the following moves.

\subsubsection{Shift move}
The shift move proposes to modify the position of one randomly chosen changepoint. The index $j$ is uniformly chosen from $\{1, \ldots, k \}$, and a new position $\tau_j^{\prime}$ is uniformly sampled from $\{\tau_{j-1}+1, \ldots, \tau_{j+1}-1 \}$. The parameters $m_j$, $m_{j+1}$ and $\gamma_{j,:}$ are not modified and we replace $\gamma_{j+1,:}$ by $\gamma_{j+1,:}^{\prime} = S^{u}(\gamma_{j+1,:})$, with $u = \tau_j^{\prime} - \tau_j$. 

As noted in Section \ref{sec:addremdata}, when the support of $f_m(\cdot \, |\, \theta)$ is bounded, the move may be rejected because the updated latent variables are unvalid: If the length of the $j$-th segment is reduced by the shift move, i.e. $\tau_j^{\prime} - \tau_j$ is negative, then it is guaranteed that $\gamma_{j,:}$ remains a valid sequence of initial latent variables but not that $\gamma_{j+1,:}^{\prime}$ is valid for the extended $(j+1)$-th segment; one the other hand, if $\tau_j^{\prime} - \tau_j$ is positive then the sequence $\gamma_{j+1,:}^{\prime}$ is valid but it must be checked that $\gamma_{j,:}$ is valid for the extended $j$-th segment.

\subsubsection{Sampling a segment parameter}
A segment $j$ is uniformly chosen amongst the $k+1$ segments, and the corresponding segment parameters are changed: Either sample the initial latent variables conditional on the order of dependence which is left unchanged; or the order of dependence is sampled such that the initial latent variables must be adapted. Here, the focus is on one segment only, and therefore we temporarily drop the segment index $j$ from the notation as in Section \ref{sec:parameterspace}, and the data observed within the segment are denoted by $x_1, \ldots, x_n$, where $n = \tau_j - \tau_{j-1}$. 

\label{sec:samplesegparam}

\paragraph{An approximation to the posterior distribution of $\theta$}

First, we consider an approximation to the posterior distribution of $\theta$ that will be useful to build proposals for $m$ and $\gamma_{1:m}$. 
In the absence of knowledge on $m$ and $\gamma_{1:m}$,  motivated by computational considerations, it is interesting to consider the posterior distribution of $\theta$ conditional on $m=0$. When the data are assumed to be exchangeable, by conjugacy of the prior for $\theta$, the posterior distribution of $\theta$ is tractable, 
\begin{align}
\hat{\pi}( \theta \, | \, x_{1:n} ) = \pi(\theta \, | \, m=0, x_{1:n} ). 
\end{align}
Based on this approximation, a natural estimator for $\theta$ is
\begin{align}
\hat{\theta} = \text{arg max}_{\theta} \,  \hat{\pi}( \theta \, | \, x_{1:n} ).
\end{align}

\paragraph{Updating $\gamma_{1:m}$ conditional on $m$}

\label{sec:updategammacondim}

The move consists in proposing $\gamma_{1:m}^\prime$ conditional on $m$ and $\gamma_{1:m}$. 
Recall from Section \ref{sec:characparameterspace} that, conditional on the observed data, the support of the initial latent variables is $\mathcal{Y}_m$, and note that $\gamma_{1:m} \in \mathcal{Y}_m$ if and only if, for all $r=1, \ldots, m$, $\gamma_r \in \mathcal{Y}_m^r$ where 
\begin{align}
\mathcal{Y}_m^r = \left\lbrace \gamma_r \in \mathcal{F} \, | \, L^{m}_{r} \leqslant  \gamma_r \leqslant U^{m} - \sum_{i \neq r} \gamma_{i} \right\rbrace,
\end{align}
if $\mathcal{F}$ is bounded and $\mathcal{Y}_m^r = \mathcal{F}$ if $\mathcal{F}$ is unbounded.

We consider two distinct scenarios based on the nature of the latent variables. If the latent variables are discrete valued, then $\gamma_{1:m}^\prime$ are proposed via Gibbs sampling. It follows from the discussion in Section \ref{sub:condlik} that, for all $r=1, \ldots, m$, the full conditional distribution of $\gamma_r$ is 
\begin{align}
\pi( \gamma_r | x_{1:n}, \gamma_{-r}, m) \propto L( x_{1:n}, \gamma_{1:m}  | m)  \mathbb{1}_{  \mathcal{Y}_m^{r} }( \gamma_r),
\end{align}
where $\gamma_{-r} = (\gamma_{1}, \ldots,\gamma_{r-1}, \gamma_{r+1}, \ldots,  \gamma_{m})$. If the support of the latent variables is continuous, then 
Gibbs sampling is not possible in general; instead, for all $r=1, \ldots, m$, sample $\gamma_r^\prime$ from
the distribution with step function density
\begin{align}
\label{eq:approxfulcond}
q(\gamma_r | x_{1:n}, \gamma_{-r}, m) \propto \sum_{i=1}^N L( \gamma_{1:m}, y_{1:n} | m) \mathbb{1}_{[\gamma^{(i)}, \gamma^{(i+1)})}( \gamma_r),
\end{align}
where $\gamma^{(1)} <  \cdots < \gamma^{(N)}$ form an equally spaced grid on the largest interval $\mathcal{Y}_m^{r*} \subseteq \mathcal{Y}_m^r$ satisfying
\begin{align}
\int_{ y_r \in \mathcal{Y}_m^{r*}} f_m(\gamma_r \, | \, \hat{\theta}) \text{d} \gamma_{r} \leqslant \eta. 
\end{align}
The greater $N>1$ and $ 0 < \eta  < 1$, the more accurate the step function approximation of the full conditional distribution of $\gamma_r^\prime$ in (\ref{eq:approxfulcond}). The tuning parameters $N$ and $\eta$ can be chosen via pilot runs investigating the trade-off between precision and computational cost.

\paragraph{Updating $m$ and $\gamma_{1:m}$}

When $m$ is replaced by some $m^{\prime}$, it is desirable that $m^{\prime} \in \mathcal{M}$, and we must propose some revised initial latent variables $\gamma^{\prime}_{1:m^{\prime}}$.

To sample $m^{\prime}$, whose full conditional distribution is not tractable in general, we consider a proposal distribution that relies on the following observations: the joint likelihood of the jumps $( \nabla x_t)$ defined in (\ref{eq:diffbdef}) is not tractable in general due the dependence of the jumps; yet, the jumps are identically distributed with mean $E[ \nabla x_t ] = 0$ and variance $V[ \nabla x_t ] = 2  g(\theta, m)$ for some function $g$ which depends on the marginal distribution  of the latent variables $f_m(\cdot \, |\, \theta)$; and therefore the approximation of the likelihood of the jumps
\begin{align}
\hat{L}( \nabla x_{1:n} | \hat{\theta}, m) = \prod_{t} \phi \left(  \nabla x_t | 0, 2  g(\hat{\theta}, m)  \right),
\end{align}
where $\phi(.| \mu, \sigma^2)$ is the density function of the normal distribution  with mean $\mu$ and variance $\sigma^2$, is tractable and depends on $m$. The proposed order of dependence $m^\prime$ is sampled from the distribution with probability mass function
\begin{align}
q(m^\prime |  \nabla x_{1:n} ) \propto \hat{L}( \nabla x_{1:n} | \hat{\theta}, m^\prime) \pi(m^\prime) \mathbb{1}_{ \mathcal{M} }( m^\prime ) .
\end{align}
Then, according to Section \ref{sec:updategammacondim},  $\gamma_{1:m^{\prime}}^\prime$ is proposed conditional on $m^\prime$ and $\gamma_{1:m^\prime}^{*}$, where
\begin{align}
\gamma_{1:m}^{*} = \text{arg max}_{ \, \gamma_{1:m^\prime}^{*} \in \mathcal{Y}_{m^\prime} }  \prod_{r} f_{m^\prime}(\gamma_r^{*} | \hat{\theta})
\end{align} 
is an estimator of the initial latent variables in $\mathcal{Y}_{m^\prime}$ which can be derived efficiently but does not take into account the dependence of the data.

\subsubsection{Death and birth moves}
If a death move is proposed, an index $j$ of one element of $\tau_{1:k}$ is uniformly chosen, and the corresponding changepoint is removed, resulting in $k^{\prime} = k-1$ changepoints with positions $\tau_{1:k^{\prime}}^{\prime} = \left(\tau_{1:(j-1)}, \tau_{(j+1):k} \right)$. The parameters corresponding to the segments which are not impacted by the move are re-indexed but kept unchanged, and it is natural to propose the parameters for the $j$-th segment resulting from $\tau_{1:k^{\prime}}^{\prime}$, namely  $( m^{\prime}_j, \gamma_{j,:}^{\prime} )$, based on the parameters of either the original $j$-th or $(j+1)$-th segment. 
Specifically, let $i$ be either the index $j$ or $j+1$ with probability proportional to the length of the segment with index $i$, and then set $m_j^{\prime}$ to $m_i$, and $\gamma_{j,:}^{\prime}$ to $S^{(\tau_j - \tau_i)}(\gamma_{i,:})$. 
Note that the death move may then be seen as the extension of one of the segments on either side of the deleted changepoint. 

The above death move may be reversed by the following birth move. Draw $\tau_j^{\prime}$ uniformly from $\{2, \ldots, T\} \setminus \tau_{1:k}$, and obtain $\tau_{1:k^{\prime}}^{\prime}$ by inserting $\tau_j^{\prime}$ into $\tau_{1:k}$ at the $j$-th position, resulting in $k^{\prime} = k + 1$ changepoints. Let $i$ be either the index $j$ or $j+1$ with probability proportional to $\tau_{i}^\prime - \tau_{i-1}^\prime$. Set $m_i^{\prime}$ to $m_j$ and $\gamma_{i,:}^{\prime}$ to $S^{(\tau_i^{\prime} - \tau_j^{\prime})}(\gamma_{j,:} )$, and finally propose the segment parameters of the new segment using the approach given in Section \ref{sec:samplesegparam}.

\subsection{Sampler initialisation}
\label{sec:samplerinit}

To speed-up the convergence of the sampler for the moving-sum changepoint model, the sample chain is initialised as follows: the changepoint parameters are set to be the changepoint estimates corresponding to the standard changepoint model; and, for each segment, the order of dependence is set to $0$. Hence, the sampler begins with a sensible positioning of the changepoints obtained at a limited computational cost.

\subsection{Changepoint estimation} 
\label{sec:changepointestime}
To give an account of the posterior distribution of changepoints, following \citet{Green1995}, it is natural to consider the posterior marginal distribution of the number of changepoints $k_i$, and the posterior distribution of the changepoint positions $\tau_{1:k}$ conditional on $k_i$. However, in practice, it will also be of interest to report a point estimate $(\hat{k}, \hat{\tau}_{1:\hat{k}})$ for the changepoint parameters $(k, \tau_{1:k})$. In this article, the point estimate $(\hat{k}, \hat{\tau}_{1:\hat{k}})$ is defined as follows: $\hat{k}$ is the MAP number of changepoints; and $\hat{\tau}_{1:\hat{k}}$ are the MAP changepoint positions of dimension $\hat{k}$.

\section{Simulation study}
\label{sec:simulatoinstudy}
This section describes a simulation study that demonstrates the benefits of the moving-sum changepoint model in comparison to the standard Bayesian changepoint model (\ref{eq:classicconiid}), which assumes the data are exchangeable within segments, and the DeCAFS model \citep{DeCAFS}, which detects abrupt changes in normal data with local fluctuations and autocorrelated noise. 

\subsection{Synthetic data}

Different scenarios were assumed to sample time series of length $T=1 \, 200$ from the moving-sum changepoint model with segment model defined in Example \ref{eg:normalgamma}: within each segment, the data are $m$-dependent and marginally normally distributed with mean $\mu$ and variance $\sigma^2$ for some segment specific parameters $m$, $\mu$ and $\sigma^2$. 
Three scenarios for the changepoint parameters were established: $k=0$ changepoint, $k=3$ changepoints with positions $\tau_{1:3}=(300, 600, 900)$, and $k=7$ changepoints with positions $\tau_{1:7}= (150, 300, 450, 600, 750, 900, 1050)$. 
The orders of dependence for each segment were sampled independently from $\text{Geometric}(\nu)$ for some $0 \leqslant \nu  \leqslant 1$. The segment mean parameters are set such that, for all $j=1, \ldots, k+1$, $\mu_j = \mu$ if $j$ is odd and $-\mu$ otherwise, for some $\mu \in \mathbb{R}$. The segment parameters $\sigma^{-2}_1, \ldots, \sigma^{-2}_{k+1}$ were sampled independently from $\text{Gamma}(\alpha_0, 100)$ for some $\alpha_0>0$. A grid of parameters such that $\nu \in \{1, 0.4, 0.3, 0.2, 0.1 \}$, $\mu \in \{1, 2, 4, 8 \}$ and  $\alpha_0 \in \{ 5, 10, 25, 50\}$ was considered for the experiments. For each scenario, $10$ simulations were performed.

\subsection{Model comparison}
\label{sec:simumodelcomp}
For each simulation, three different models were used to infer changepoint estimates from the data: the moving-sum changepoint model, the standard changepoint model (\ref{eq:classicconiid}), and the DeCAFS model \citep{DeCAFS}. For the moving-sum changepoint model, the data within segments are assumed to follow the model given in Example \ref{eg:normalgamma} with $\alpha = \alpha_0$, $\beta = 100$ and $\lambda = 0.05 \beta / \alpha$; and the prior for the orders of dependence is assumed to be $\text{Geometric}(0.15)$.
Ten independent sample chains of size $20 \, 000$, after a burn-in of size $5 \, 000$, were obtained via the MCMC algorithm described in Section \ref{sec:rjmcmc} with tuning parameters $N=100$ and $\eta=0.99$ for the proposal of the initial latent variables (\ref{eq:approxfulcond}). Changepoint estimates obtained from each independent sample chain, as described in Section \ref{sec:changepointestime}, will be compared to assess the convergence of the sampler. 
Moreover, for the standard Bayesian changepoint model, changepoint estimates are obtained as for the moving-sum changepoint model with the only difference that, for each segment, we fix the order of dependence $m=0$, ensuring the data are assumed to be exchangeable. Finally, for the DeCAFS model, changepoints were estimated with the default implementation described in \citet{DeCAFS}.

To compare changepoint estimations, we use the $F1$ score. 
A changepoint $\tau$ is said to be detected if there is an estimated changepoint $\hat{\tau}$ such that $|\tau - \hat{\tau}| \leqslant \epsilon$ for some error tolerance of size $\epsilon = 5$.  Given some changepoints, 
the $F1$ score of the changepoint estimates is 
\begin{align}
F1 = \frac{2 P R}{ P + R}  \in [0, 1],
\end{align}
where $R$ and $P$ denote the recall, the proportion of simulated changes that are detected, and the precision, the proportion of detected changes which are correct, respectively. The greater the $F1$ score, the better the estimation.  
To compare the three changepoint models of interest, for each simulation the $F1$ score was computed for each estimation of the simulated changepoints $(k, \tau_{1:k})$.

\begin{figure}[t!]
\centering
\includegraphics[height=0.48\textwidth]{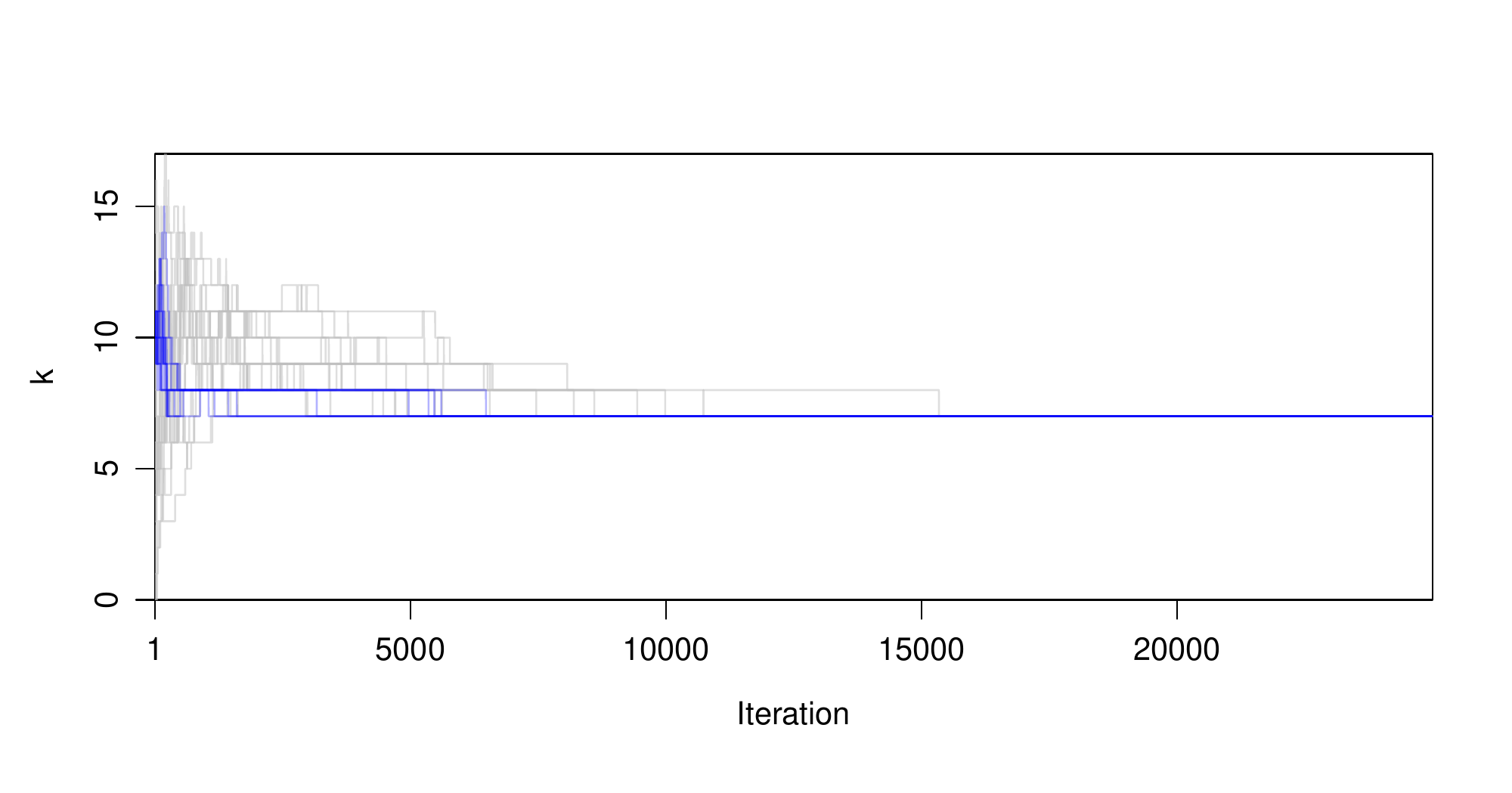}
\caption{Thirty independent sample chains for the number of changepoints, obtained via the MCMC algorithm for the moving-sum changepoint model, for one selected time series with $k=7$. In blue: ten sample chains initialised with the proposed initialisation strategy. In grey: twenty sample chains initialised with randomly selected changepoint parameters such that $0 \leqslant k \leqslant 20$. }
\label{fig:simuindechains}
\end{figure}

\subsection{Sampler convergence}
Both for the moving-sum changepoint model and the standard changepoint model, for each scenario, the variance of the F1 scores of changepoint estimates obtained from ten independent sample chains was less than $0.1$, suggesting the samplers converge. 

Moreover, we further assess the convergence of the sampler for one randomly selected simulation with $k=7$. 
Following Section \ref{sec:samplerinit}, the ten independent sample chains for the moving-sum changepoint model were initialised with the changepoint estimates corresponding to the standard changepoint model. For comparison purposes only, twenty extra independent sample chains, initialised with randomly selected changepoint parameters such that $0 \leqslant k \leqslant 20$, were computed for the moving-sum model, as described in Section \ref{sec:simumodelcomp}. Figure \ref{fig:simuindechains} displays the thirty independent sample chains for the number of changepoints $k$. All sample chains converge to the same number of changepoints, namely $k=7$, and the ten sample chains corresponding to our proposed initialisation, indicated in blue, converge faster, illustrating the benefits of our proposed sampling strategy.

\begin{figure}[t!]
\centering
\includegraphics[height=0.7\textwidth]{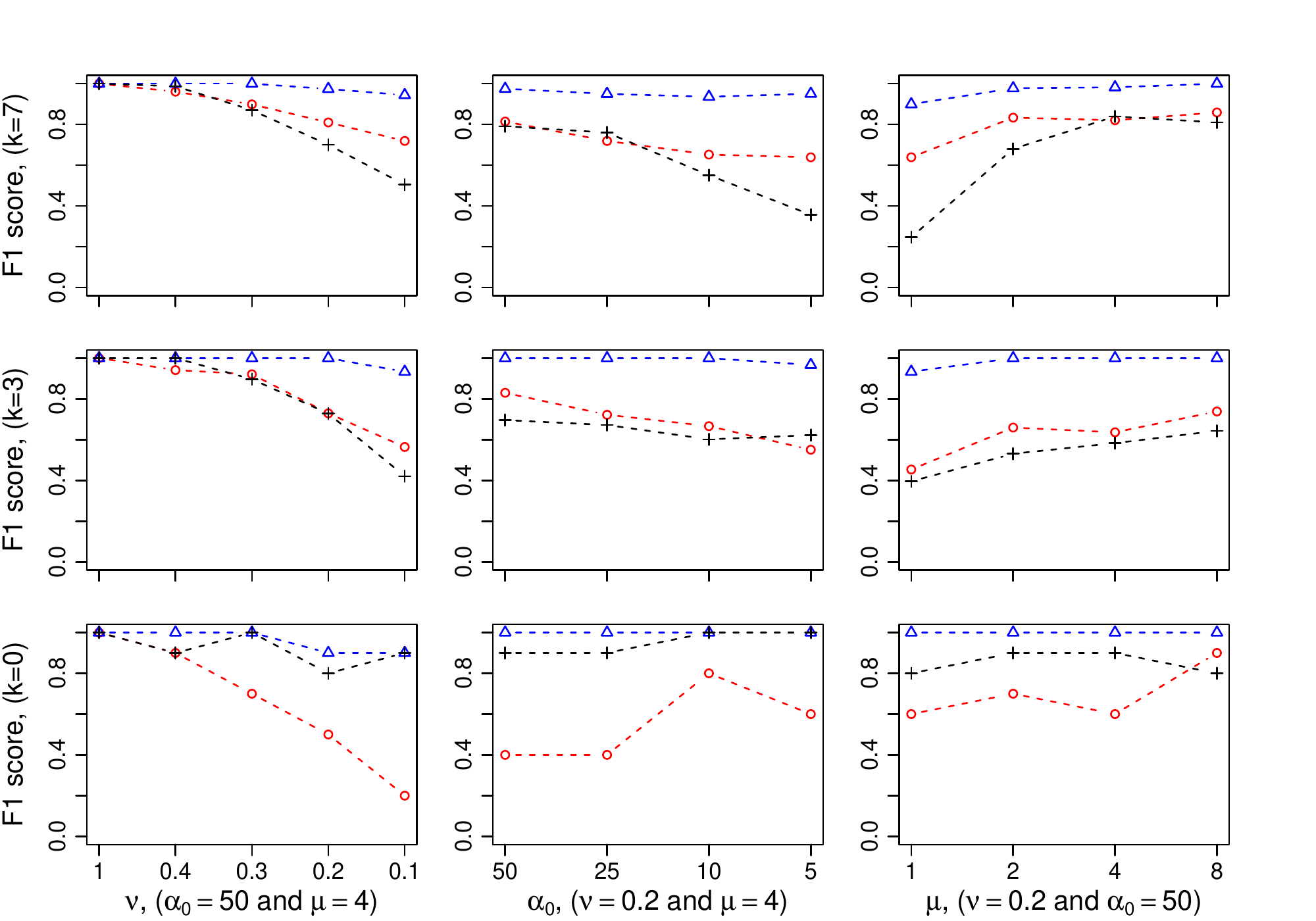}
\caption{$F1$ score for the three changepoint models of interest (Moving-sum: blue triangles. Standard: red circles. DeCAFS: black crosses) for each scenario considered in the study. 
 }
\label{fig:f1scores}
\end{figure}

\subsection{Results}

Figure \ref{fig:f1scores} displays the average $F1$ score of the three changepoint models of interest for each scenario considered in the study. 
For each scenario, the moving-sum model, with a $F1$ score close to $1$, outperforms both the standard changepoint model and the DeCAFS model. 
As $\nu$ decreases, the orders of dependence $m_1, \ldots, m_{k+1}$ tend to increase, resulting in lower $F1$ scores for both the standard changepoint model and the DeCAFS model. 
As $\sigma_0$ increases, the data are noisier, and therefore the performance of the standard model and the DeCAFS model decreases. As $\mu$ decreases, the changes in the mean of the data are smaller, leading to lower $F1$ scores for both the standard model and the DeCAFS model. Moreover, the DeCAFS model's performance is relatively good when $k=0$, but it tends to deteriorate as the number of segments increases. 

The simulation study demonstrates the benefits of the proposed moving-sum changepoint model, which is suitable for $m$-dependent data within segments, in contrast to the two other changepoint models of interest.  The standard changepoint model cannot capture temporal dynamics within segments, and consequently the proposed segmentations of the simulated data are not satisfactory, except when $\nu=1$, that is when the data are exchangeable within segments. Moreover, the DeCAFS model admits the data may not be exchangeable within segments; yet 
the model assumes the dependence structure of the data is the same for each segment; therefore, the DeCAFS model is not appropriate when we have multiple segments with distinct orders of dependence.

\section{Applications}
\label{sec:applications}
Two applications are considered to demonstrate the benefits of the proposed changepoint model for non-exchangeable data: computer network monitoring via change detection in count data, and detection of breaks in daily prices of a stock.

For each application, the moving-sum changepoint model is compared with the standard changepoint model for exchangeable data within segments (\ref{eq:classicconiid}). 
For each model, changepoint estimates are derived, as described in Section \ref{sec:changepointestime}, from a  
sample of size $50 \, 000$ 
obtained via the MCMC algorithm proposed in Section \ref{sec:rjmcmc}, with a burn-in of $10 \, 000$ iterations. 
Moreover, since DeCAFS is not suitable for count data \citep{DeCAFS}, the moving-sum model is compared with DeCAFS for the daily stock prices only.

\subsection{Change detection in enterprise-wide computer network traffic}
\label{sec:cyberapp}
A cyber attack typically changes the behaviour of the target network. Therefore, to detect the presence of a network intrusion, it can be informative to monitor for changes in   
computer network traffic.

\citet{unifieddata} presents a data set summarising $90$ days of network events collected from the Los Alamos National Laboratory enterprise network, which is available online at \url{http://lanl.ma.ic.ac.uk/data/2017/}. Each recorded network event gives the start time and the duration of a transfer of packets from a network device to another. In addition, a destination port is associated to each network event, which describes the purpose of the transfer of packets: for example, web, email, remote login or file transfer. For the purpose of this article, events that do not correspond to the $100$ most recurrent destination ports in the data were discarded, thereby restricting the analysis to the most common network activities.  

It can be informative to monitor for temporal changes in counts of 
network events. For demonstration purposes, we consider the data $x_{1}, \ldots, x_{T}$ where, for all $t$, $x_t$ denotes the number of network events that are in progress across the network during the $t$-th second between 10:00 and 10:20 on day $22$ of the data collection period. The data are displayed in Figure \ref{fig:Netflow}. 
It is of interest to detect temporal changes in the distribution of $x_{1}, \ldots, x_{T}$.

Two models are used to estimate changepoints for the network data: the moving-sum and the standard changepoint model. For each changepoint model, the segment model for negative binomial data defined in Example \ref{eg:negbin} is assumed with $\alpha=3$, $\beta=1$ and $r=3000$. For each segment, $m \sim \text{Geometric}(0.1)$ for the moving-sum model, and $m=0$ for the standard changepoint model. 

Figure \ref{fig:Netflow} displays the changepoint estimates for each model, and the MAP segment orders of dependence for the moving-sum changepoint model. 
Both changepoint models detect clear discontinuities, such as the ones observed near the 420th and 880th seconds, which may be evidence for malicious activity on the network. However, small fluctuations and local temporal correlations, such as the ones between the 100th and 380th seconds, which correspond to normal temporal dynamics of the network behaviour, give rise to changepoints for the standard model but not the moving-sum changepoint model. Hence, the proposed changepoint model results in a segmentation of the data that is more adapted to network monitoring. 

\begin{figure}[t!]
\centering
\includegraphics[height=0.9\textwidth]{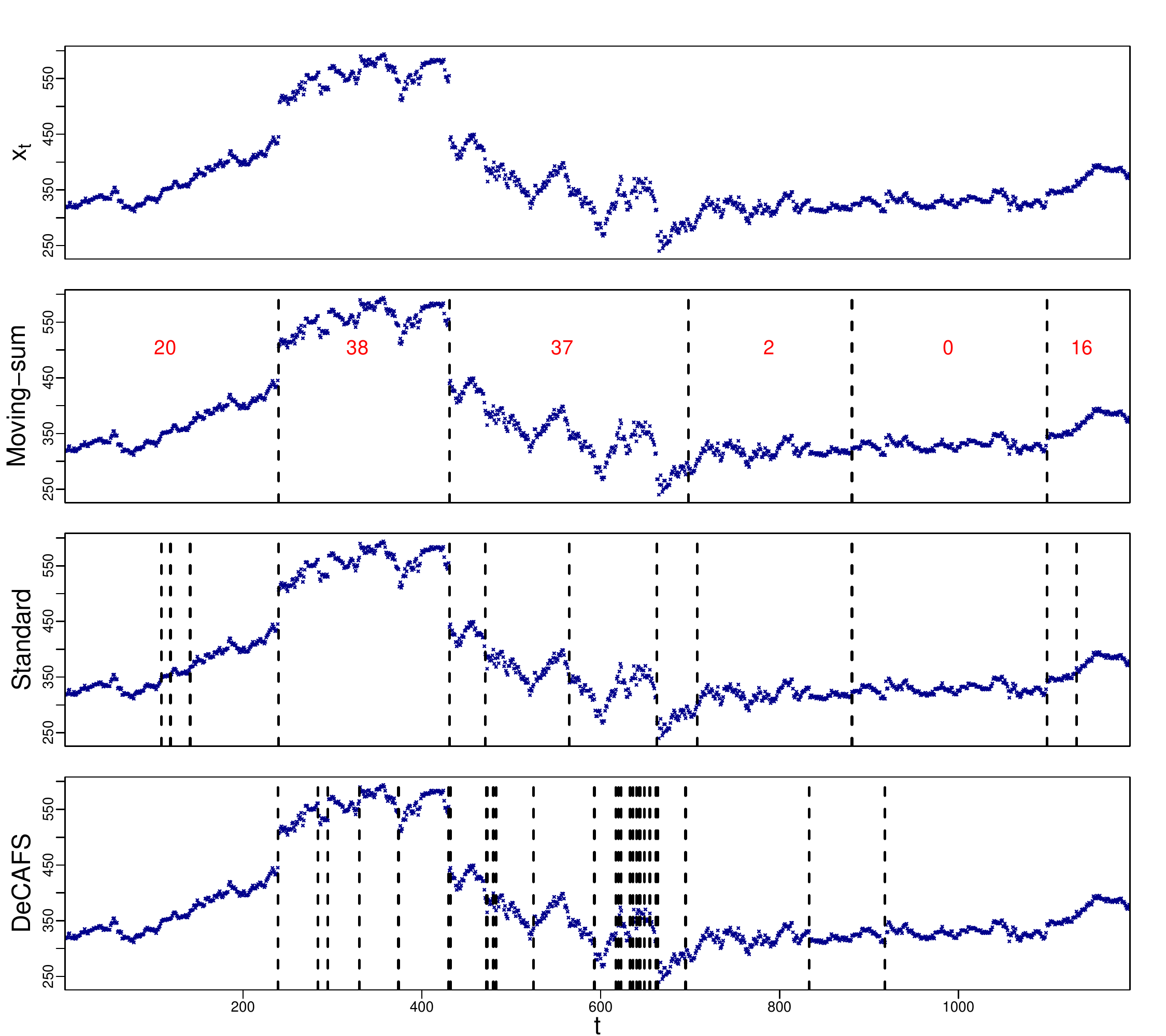}
\caption{Daily prices of the stock SBRY.L from $20^{\text{th}}$ February 2006 to $16^{\text{th}}$ November 2010. Vertical lines indicate estimated changepoints for three models: moving-sum, standard and DeCAFS changepoint models. Numbers in red indicate the MAP order of dependence $m$ for each segment for the moving-sum changepoint model. }
\label{fig:Stock}
\end{figure}

\subsection{Change detection in financial time series}
For economists and investors, it can be of interest to detect changepoints in financial time series, such as daily prices of a stock. 
Changes can be monitored to assess  
the impact of economic policies, or can indicate shifts in market behaviours. It is often reductive to assume the data are exchangeable within segments. 

For demonstration purposes, this article considers  the price of the stock Sainsbury plc (SBRY.L). For all $t$, let $x_t$ be the closing price of the stock SBRY.L at the $t$-th day between $20^{\text{th}}$ February 2006 to $16^{\text{th}}$ November 2010. 
The data $x_1, \ldots, x_{T}$, which are available online at \url{finance.yahoo.com}, are displayed in Figure \ref{fig:Stock}.

Three models are considered for the data: the moving-sum changepoint model, the standard changepoint model, and DeCAFS using the default implementation described in \citet{DeCAFS}. For the moving-sum and the standard changepoint models, the segment model for normal data defined in Example \ref{eg:normalgamma} is assumed with $\alpha=7$, $\beta=1$,  $\mu_0=350$ and $\lambda=1$. For each segment,  $m \sim \text{Geometric}(0.1)$ for the moving-sum model, and $m=0$ for the standard changepoint model.

Figure \ref{fig:Stock} displays the changepoint estimates for each model of interest. 
In contrast with the standard changepoint model, the moving-sum changepoint model captures temporal dynamics of the stock price, and therefore market trends are not unnecessarily segmented. For example, smooth drifts of the stock price, such as the ones between the 50th and 150th days and between the 450th and 700th days, give rise to multiple changepoints for the standard model but not for the moving-sum model. Moreover, the proposed model detects changes in the mean  as well as changes in the level of dependency of the data, so that it detects both shifts in price levels and changes in the temporal dynamics of prices, which are both of interest to financial analysts. The estimated orders of dependence for the moving-sum model vary greatly across segments, suggesting prices have been subject to distinct market dynamics. Although DeCAFS admits the data may be non-exchangeable within segments, the dependence structure is assumed to be the same for each segment. As a result, DeCAFS is not suitable to segment these data where the dependence structure changes greatly from one market trend to another.

\section*{Code}

The \textit{python} code used for this work is available in the \textit{GitHub} repository \href{https://github.com/karl-hallgren/mvsum}{\texttt{karl-hallgren/mvsum}}.

\section*{Acknowledgements}
The authors gratefully acknowledge funding from EPSRC.

\bibliography{bibli.bib}

\clearpage 

\appendix

\section*{Appendices}

\section{Proofs}

\subsection{Proof of Proposition \ref{prop:mdjump}}
\label{appen:proofs}

\begin{itemize}
\item[(i)] For all $n$ and for all reals $z_1, z_2, \ldots, z_n$,
\begin{align}
\label{eq:min1}
\min\{ z_1, \ldots, z_n \} = -\max\{ -z_1, \ldots, -z_n  \},
\end{align}
and therefore $U^{m} = x_1 - L^{m}_{m+1}$ for all $m$ by definition (\ref{eq:yinitbounds}). 

Moreover, for all $\ell$ and for all reals $(z_{ij})$,
\begin{align*}
\sum_{i = 1}^{\ell} \max \{0, z_{i1}, z_{i2}, z_{i3}, \ldots \} \geqslant \max \left\lbrace 0, z_{11}, z_{11} + z_{21}, z_{11} + z_{21} + z_{31}, \ldots, \sum_j \sum_{i=1}^{\ell} z_{ij} \right\rbrace,
\end{align*}
and therefore, with $\ell$ such that $ m+1 = \ell(\dd+1)$ and using the notations introduced in (\ref{eq:yinitbounds}), for all $r=1, \ldots, \dd+1$,
\begin{align}
\label{eq:ineqproof}
\sum_{j=0}^{\ell-1} L^{m}_{j(\dd+1)+r} &= \sum_{j=0}^{\ell-1} \max \left\lbrace 0, \difff x_{j(\dd+1)+r} , \, \ldots \, , \sum_{q = 0 }^{\kappa_{j\mn^{\prime}+r}}  \difff x_{(q\ell+j)(\dd+1)+r}  \right\rbrace \nonumber \\
& \geqslant L^{\dd}_r.
\end{align}

 As a result, for all $\gamma_{1:m}$ such that $U^{m} \geqslant \sum_r \gamma_r$ and $\gamma_r \geqslant L^{m}_r$ for all $r=1, \ldots, m$, that is $\gamma_{1:m} \in \mathcal{Y}_m$, it follows that
\vspace{-4mm}
\begin{align*}
U^{\dd} = x_1 - L^{\dd}_{\dd+1}  &\geqslant x_1 - \sum_{j=0}^{\ell-1} L^{m}_{j(\dd+1) + (\dd+1)}  \\ 
&=  U^{m} - \sum_{j=0}^{\ell-2} L^{m}_{(j+1)(\dd+1)} \\
 & \geqslant \left( \sum_{r=1}^{m} \gamma_r \right) - \sum_{j=0}^{\ell-2} \gamma_{(j+1)(\dd+1)} = \sum_{r=1}^{\dd} \gamma^{\prime}_r
\end{align*}
and  $\gamma^{\prime}_r \geqslant \sum_j L^{m}_{j(\dd+1) + r} \geqslant L^{\dd}_r$ for all $r= 1, \ldots, \dd$, that is $\gamma^{\prime}_{1:\dd} \in  \mathcal{Y}_{\dd}$.
\item[(ii)] Follows from (i).
\item[(iii)] It is immediate from the definition of $J_{\dd}$ that the transformation between $\gamma_{1:m}$ and $( \gamma^{\prime}_{1:\dd}, \gamma_{(\dd+1):m}  )$ is one-to-one and its Jacobian is equal to 1. Now, for all $\gamma^{\prime}_{1:\dd} \in J_{\dd}(\mathcal{Y}_m)$, it is of interest to characterise the set of $\gamma_{(\dd+1):m}$ such that $(\gamma_{1:\dd}, \gamma_{(\dd+1):m}) \in \mathcal{Y}_m$, with 
\begin{align*}
\gamma_r = \gamma^{\prime}_r -\sum_{j=1}^{\ell-1} \gamma_{j(\dd+1)+r}
\end{align*}
for all $r=1, \ldots, \dd$.  Recall that $\mathcal{Y}_m = \mathds{Y}[U^{m}, L^{m}_{1:m}]$, where $\mathds{Y}$ is defined in (\ref{eq:mathcaldef}), and observe that
\begin{align*}
\gamma^{\prime}_r - L^{m}_r \geqslant \gamma^{\prime}_r - \gamma_r =  \sum_{j=1}^{\ell-1} \gamma_{j(\dd+1)+r}
\end{align*}
for all $r=1, \ldots, \dd$. As a result, the set of interest consists of the $\gamma_{(\dd+1):m}$ such that 
\begin{align*}
\left( \gamma_{j(\dd+1) + r} \right)_{j>0} \in \mathds{Y}\left[\gamma^{\prime}_r - L^{m}_r, \, ( L^{m}_{j(\dd+1) + r}  )_{j>0} \right]
\end{align*}
for all $r=1, \ldots, \dd$ and
\begin{align*}
\left( \gamma_{j(\dd+1)} \right)_{j>1} \in \mathds{Y} \left[U^{m} - \sum_r \gamma^{\prime}_r, \, (L^{m}_{j(\dd+1)})_{j>1}  \right].
\end{align*}
\end{itemize}%

\subsection{Proof of Proposition \ref{prop:almostsureconvergence}}
\label{sec:apendproofpropconvergence}
%
Note that (\ref{eq:diffbdef}) and (\ref{eq:diffequation}) imply that $\difff x_t = - \diffb^{(m+1)} y_{t+1}$ for all $t$. Hence, by definition of the bounds in (\ref{eq:yinitbounds}), for all $n$, 
\begin{align}
\label{eq:proofconvebound}
L^{\dd}_r(x_{1:n}) &= \max \left\lbrace 0, -\diffb^{(m+1)} y_{r+1} , \ldots, -\sum\limits_{q = 0 }^{\kappa_{r}(n)}  \diffb^{(m+1)} y_{q(\dd+1)+r+1}  \right\rbrace ,
\end{align}
with $\kappa_r(n)$ defined to be the largest $q\in \mathbb{N}_{0}$ such that $q(\dd+1)+r+1 \leqslant n$, for all $r$. 

If $\dd \in \mathcal{D}(m)$, that is there is some $\ell>0$ such that $m+1 = \ell(\dd+1)$, then for all $q$, 
\begin{align*}
\diffb^{(m+1)} y_{q(\dd+1)+r+1}  = y_{q(\dd+1)+r+1} - y_{ (q-\ell)(\dd+1)+r+1}, 
\end{align*}
and we note that $ q(\dd+1) + r + 1= (q + \ell - \ell )(\dd+1) + r + 1$. Therefore
\begin{align*}
L^{\dd}_r(x_{1:n}) &= \left( \sum_{q=0}^{\ell-1} y_{(q-\ell)(\dd+1) + r + 1} \right) - B_n = \gamma^{\prime}_r - B_n,
\end{align*}
where
\begin{align*}
B_n= \min \left\lbrace   \sum_{q=j-\ell}^{j-1} y_{q(\dd+1) + r + 1} , \quad j = 0, \ldots, \kappa_r(n) \right\rbrace
\end{align*}
is a random variable which converges almost surely to $0$, since it is non-increasing and converges in probability to the infimum of the support of $f_m(\, \cdot \, | \theta)$ which is $0$. Hence $L^{\dd}_r(x_{1:n})$ converges almost surely to $\gamma^{\prime}_r $.

However if $\dd \notin \mathcal{D}(m)$ then for all $q_1, q_2 \in \{0, \ldots, \kappa_r(n) \} $, we have that $q_1(\dd+1)+r+1 = q_2(\dd+1)+r+1$ if and only if $q_1 = q_2$, so that no cancellations occur in (\ref{eq:proofconvebound}). Hence $L^{\dd}_r(x_{1:n})$, which is non-decreasing, converges almost surely to the supremum of the support of $f_m(\, \cdot \, | \theta)$ which is $\infty$.

Finally, the results on the upper bound then follow from the observation that 
\begin{align*}
U^{\dd}(x_{1:n}) &= x_1 - L^{\dd}_{\dd+1}(x_{1:n}) \\
&= \left( \sum_{r=1}^{\dd} \gamma^{\prime}_r \right) + \left( \sum_{q=0}^{\ell-1} y_{(q-\ell + 1)(\dd+1) + 1} \right)  - L^{\dd}_{\dd+1}(x_{1:n}).
\end{align*}

\end{document}